\documentclass[aps,prd,showpacs,nofootinbib,superscriptaddress,preprint,tightenlines,eqsecnum,longbibliography]{revtex4}
\usepackage{amssymb,amsmath}
\usepackage{graphicx,enumerate}

\usepackage{subfigure}

\usepackage{leftidx}

\newcommand{\vd}[2]{\frac{\delta #1}{\delta #2}}   

\newcommand{\md}{{\mathrm d}}

\newcommand{\ct}{c_{(t)}}

\newcommand{\pbi}[1]{{\underset{^\leftarrow}{{#1}}}} 
\newcommand{\WIHeq}{\overset{_\Delta}{=}}           

\newcommand{\A}{\mathcal{A}}     

\newcommand{\W}{\mathbb{W}}
\newcommand{\V}{\mathbb{V}}
\newcommand{\U}{\mathbb{U}}
\newcommand{\Y}{\mathbb{Y}}

\newcommand{\om}{{\rm o}}   

\def\de{\delta}
\def\k{\kappa}

\begin{document}

\title{Weakly Isolated Horizons: \\
First order actions and gauge symmetries}
\author{Alejandro Corichi}
\email{corichi@matmor.unam.mx}
\affiliation{Centro de Ciencias Matem\'aticas, Universidad Nacional Aut\'onoma de
M\'exico, UNAM-Campus Morelia, A. Postal 61-3, Morelia, Michoac\'an 58090,
Mexico}
\affiliation{Center for Fundamental Theory, Institute for Gravitation and the Cosmos,
Pennsylvania State University, University Park
PA 16802, USA}
\author{Juan D. Reyes}
\email{jdreyes@uach.mx, jdrp75@gmail.com}
\affiliation{Facultad de Ingenier\'\i a,
Universidad Aut\'onoma de Chihuahua, 
Nuevo Campus Universitario, Chihuahua 31125, Mexico}
\affiliation{Centro de Ciencias Matem\'aticas, Universidad Nacional Aut\'onoma de
M\'exico, UNAM-Campus Morelia, A. Postal 61-3, Morelia, Michoac\'an 58090,
Mexico}
\author{Tatjana Vuka\v{s}inac}
\email{tatjana@umich.mx}
\affiliation{Facultad de Ingenier\'ia Civil, Universidad Michoacana de San Nicol\'as de Hidalgo, 
Morelia, Michoac\'an 58000, Mexico}

\begin{abstract}
The notion of Isolated Horizons has played an important role in gravitational physics, being useful from the characterization of the endpoint of black hole mergers to (quantum) black hole entropy. 
With an eye towards a canonical formulation we consider general relativity in terms of connection and vierbein variables and their corresponding first order actions. We focus on two main issues: i) The role of the internal gauge freedom that exists, in the consistent formulations of the action principle, and ii) the role that a 3+1 canonical decomposition has in the allowed internal gauge freedom. 
More concretely, we clarify in detail how the requirement of having well posed variational principles compatible with general {\it weakly} isolated horizons (WIHs) as internal boundaries  does lead to a partial gauge fixing in the first order descriptions used previously in the literature. We consider the
standard Hilbert-Palatini action together with the Holst extension (needed for a consistent 3+1 decomposition), with and without boundary terms at the horizon. We show in detail that, for
the complete configuration space --with no gauge fixing--, while
the Palatini action is differentiable without additional surface terms at the inner WIH boundary, the more general Holst action is not. The introduction of a surface term at the horizon --that renders the action for asymptotically flat configurations differentiable-- does make the Holst action differentiable, but only if one restricts the configuration space and partially reduces the internal Lorentz gauge. 
For the second issue at hand, we show that upon performing a 3+1 decomposition and imposing the time gauge, there is a further gauge reduction of the Hamiltonian theory in terms of Ashtekar-Barbero variables to a $U(1)$-gauge theory on the horizon. We also extend our analysis to the more restricted boundary conditions of (strongly) isolated horizons as inner boundary.
We show that even when the Holst action is indeed differentiable without the need of additional surface terms or any gauge fixing for  Type I spherically symmetric (strongly) isolated horizons --and a preferred foliation--, this result does not go through for more general isolated or weakly isolated horizons.
Our results represent the first comprehensive study of these issues and clarify some contradictory statements found in the literature.

\end{abstract}

\pacs{04.20.Fy, 04.70.Bw, 04.20.Cv}
\maketitle
\tableofcontents

\section{Introduction}

The notion of  weakly isolated horizons (WIHs) is a quasilocal generalization of event horizons encompassing cosmological horizons and modeling black holes in quasi equilibrium \cite{PRL,AFK}. They have been used in a variety of situations, from asymptotic final states of merging black holes, to black hole entropy  in loop quantum gravity. 
As such, they have also been shown to satisfy relations that extend the validity of the zeroth and first laws of black hole event horizons.
Furthermore, a particular, highly symmetric type of horizon dubbed Type I within the hierarchy defined by WIHs, has played a key role in state counting for black hole entropy from loop quantum gravity  \cite{ABCKprl, ENPprl}.

Loop quantum gravity (LQG) \cite{ThiemannBook,AL} is rooted on a classical description of general relativity formulated as a 
first order theory. In particular it can be derived from the so called Holst action as a starting point. This action has as fundamental 
variables an $SO(3,1)$ connection ${\mathcal{A}_{\mu\, IJ}}$ and a tetrad $e_\mu^I$, and it represents an extension of the  Palatini-Hilbert action.
The canonical description of the theory is obtained after an appropriate 3+1 decomposition of the action, together with a gauge 
reduction that allows for the local gauge group to descend from $SO(3,1)$ to $SO(3)$ \cite{holst,barros}.
On the other hand, systematic analysis of action principles and their corresponding {\it covariant} Hamiltonian formulation of 
WIHs have made use of first order formulations and actions with explicit internal Lorentz gauge symmetry 
\cite{AFK, ChatterjeeGhosh}. Within this formulation, it has been shown that generalizations of the laws of black 
hole mechanics are also valid in this case. Such analyses based on the Palatini and Holst actions only make use of the covariant action and do not imply a 3+1 decomposition.

Historically, since they were partly motivated or conceived for entropy calculations within LQG, even the first definitions of spherically symmetric Type I isolated horizons relied on a first order description \cite{ABCKprl, ABF0, ABF, ACKclassical}.
This has been the source of much confusion which we aim to clarify here.
The definition of WIHs as special types of null hypersurfaces foliated by marginally trapped surfaces is purely geometrical. Whereas admitting a weakly isolated horizon certainly imposes restrictions on a spacetime $(\mathcal{M},g_{\mu\nu})$, the modern definition of WIHs does not involve any internal gauge description of the gravitational field. In principle, imposing WIH boundary conditions should not restrict the internal Lorentz gauge freedom of a first order formulation. However, as we shall argue in this work, it is the requirement of the well posedness of the first order variational principles involved  that demands a gauge reduction in the formalism.
Specifically, the technical but key point to such well posedness of the variational principle is the differentiability of the action.

A rigorous or direct derivation of the relevant gravitational degrees of freedom to quantize in terms of Ashtekar-Barbero variables 
used in LQG, as well as a canonical derivation of the first law for WIHs, should  begin with a well posed action and variational 
principle. As a second step one performs a 3+1 decomposition of the theory before quantising.
As mentioned before, we consider here the first order covariant Holst action \cite{Hojman,holst} based on orthonormal tetrads and a 
Lorentz connection. There is a good reason one should focus on the Holst action as the starting point to derive the canonical theory describing isolated horizons, for subsequent black hole entropy calculations in LQG. The Holst action is the simplest first order action whose 3+1 decomposition plus time gauge fixing leads directly to a Hamiltonian formulation in terms of Ashtekar-Barbero variables for compact and asymptotically flat spacetimes. 
The celebrated calculations of black hole entropy, however, start from the self-dual action based on complex variables, and as we have noted use the more restricted Type I isolated horizons. At present, there is no rigorous derivation of an effective theory that describes the relevant degrees of freedom living at the horizon, starting from the real Holst action.
Standard analyses for Type I  \cite{ABCKprl, ENPprl} and derivations for less symmetric horizons \cite{others} raise several questions and also add to the confusion of 
whether a U(1) or a SU(2) effective theory should emerge at the given horizon.

The relation between action principles and the necessity of including a boundary term at the horizon has not been free of controversy. For instance,
there have been some proposals for a boundary term on a general WIH horizon, needed in order to render the Holst action differentiable.  
In \cite{AFK}, it was shown that no boundary term at a WIH horizon is required for the Palatini action, however, the need for 
a surface term was pointed out already in \cite{ABF} for the self-dual action, and in \cite{ChatterjeeGhosh} and \cite{CRGV2016}  
for the general Holst action.  It turns out that the same expression used at spatial infinity for the Holst action can be used at 
a WIH horizon to write a boundary term that leads to a differentiable action and a well defined covariant formulation \cite{CorichiWilsonEwing}, but in the appropriately 
restricted configuration space \cite{CRVwih2}. However, to date no analysis of the consistency of all possible actions --with and without boundary terms-- 
is available without a partial gauge fixing for generic WIH.
There have been other proposals for boundary terms in the vielbein formalism that are gauge invariant 
\cite{Obukhov, BianchiWieland, Bodendorfer2013}. 
However, those terms involve expressions that can only be understood on null surfaces via a limiting process (for proposals in the second order, metric formulation see 
\cite{padmadhavan}). 
To our knowledge, there is no extension of the definition of those terms for null boundaries. Furthermore, those terms are ill-defined for 
asymptotically flat boundary conditions, so are not suitable to be analysed in detail and shall not be considered here further.

One issue that has not received proper attention in the literature is the connection between differentiability of the action and symmetry reductions of the internal
gauge symmetry present in the first order formulation.  Let us be precise. In the first treatments of the subject, within the restricted setting of strongly isolated 
horizons (and spherical symmetry), the action was taken to be self-dual, with an SL(2,C) internal symmetry. A gauge fixing reduced the symmetry to  U(1), which was taken 
then as the symmetry of the horizon \cite{ACKclassical,ABF}, but the necessity of such reduction was not studied. Later on, it was shown  that for this restricted set of 
horizons one could indeed maintain a larger gauge group that led, in the canonical theory, to a SU(2) symmetry \cite{ENPprl}. When the more general
definition of weakly isolated horizons was introduced, there were two restricting assumptions made \cite{AFK}. First, it was only for the Palatini action that the covariant 
Hamiltonian analysis was performed and, second, a gauge fixing of the internal symmetry was assumed at the outset. Even when this assumption was chosen for convenience, there 
was no proof of the necessity of this gauge reduction for the consistency of the formalism. In this manuscript we address these two loose ends.
Our analysis of the Holst action includes as special cases the Palatini and self-dual actions used previously in the literature. 
So this exposition will also allow us to make contact with previous treatments and clarify in detail how the requirement of having a well posed variational principle 
compatible with general WIHs as internal boundaries does lead to a partial gauge fixing in the self-dual case too.
Our work represents a first but important step towards a more direct derivation starting from the Holst action, of the effective canonical theory in terms of real 
Ashtekar-Barbero variables on more general isolated horizons.  

During our previous discussion we have mentioned that several of the previous works analyse the canonical description of the theory with an isolated horizon as an 
internal boundary. But, to be precise, to our knowledge there has been no exhaustive treatment that starts from an action, with suitable horizon terms, and performs 
a 3+1 decomposition of the theory. Such study should yield a proper understanding of the role that differentiability plays, and the way a symmetry reduction of the theory 
takes place. In this work we try to address this issue heads on. We shall not complete the task of performing a full 3+1 decomposition, but shall start such program by asking 
simpler questions: What is the role that a  3+1 decomposition will have in the geometric conditions that define a WIH, how will that affect the geometric data at the horizon and, 
what will be the  impact on the gauge symmetries allowed in the first order formulation? In the second part of this manuscript we give concrete answers to these questions.



The paper is organized as follows.  In section \ref{S_preliminaries} we recall and lay down the hierarchy of isolated horizons. To make the paper as self contained as possible, we review in detail all the relevant geometric structures and symmetries involved in describing and characterizing the different layers of the hierarchy: from general null hypersurfaces to more specialized strongly isolated horizons of Type I. This exposition will also serve to highlight the purely geometric nature of isolated horizons. We include a discussion of the freely specifiable data on isolated horizons leading to a proof of the essential uniqueness of Type I strongly isolated horizons with preferred foliations and which is key to the results in \cite{ENPprl, EngleNouiPerezPranzetti}.

In section \ref{S_FirstOrder} we review the first order formulation of general relativity in terms of the tetrad $e$ and Lorentz connection $\mathcal{A}$.
We pinpoint the gauge reduction to $\mathbb{R}^+_{\text{global}}\rtimes ISO(2)$ on a WIH boundary of previous covariant treatments using the Palatini and Holst actions. Our analysis clarifies and generalizes former claims  in \cite{BasuChatterjeeGhosh} for the local symmetries of non-expanding horizons by properly connecting them to the well posedness of the first order variational principle.

In section \ref{S_actions} we consider for the first time the differentiability of the Holst action in 
 the full configuration space describing --with no gauge fixing--  asymptotically flat spacetimes admitting a WIH as an internal boundary. We extend the proof in \cite{AFK} to this larger configuration space and show that the Palatini action is differentiable without additional surface terms at the inner WIH boundary but we confirm that the general Holst action is {\it not} differentiable by expanding and refining the arguments in \cite{ChatterjeeGhosh, CRGV2016}. An extension to the internal WIH boundary of the surface term at spatial infinity does make the Holst action differentiable but only if one restricts the configuration space and partially reduces the internal Lorentz gauge to $\mathbb{R}^+_{\text{global}}\rtimes ISO(2)$.  Finally, we prove that in the very special case where one restricts the configuration space to Type I spherically symmetric (strongly) isolated horizons and a preferred foliation, the Holst action is differentiable without the need of additional surface terms or any gauge fixing. This is consistent and  gives further support to results in  \cite{ENPprl, EngleNouiPerezPranzetti} which make use of the self-dual action.
 
In section \ref{S_decomposition} we argue how a subsequent 3+1 decomposition and time gauge fixing of the Holst action with boundary term and restricted $\mathbb{R}^+_{\text{global}}\rtimes ISO(2)$ gauge freedom
 necessarily implies an effective $U(1)$-gauge theory on the internal WIH boundary.
 We also show how demanding compatibility with a 3+1 foliation independently imposes a gauge reduction of the theory to  
 $\mathbb{R}^+_{\text{global}}\rtimes U(1)$. This gauge fixing was imposed in the original treatments of Type I horizons \cite{ABCKprl, ABF0, ABF, ACKclassical} which used the self-dual action with spinorial variables and a preferred foliation.
 
 In section \ref{S_outlook} we summarize and comment on some implications of our analysis.  
Finally Appendix \ref{A_NP} contains details of our calculations while Appendix  \ref{A_Spinors} makes direct contact with the spinorial formulation.


\section{Preliminaries}   \label{S_preliminaries}

Throughout this work we will assume spacetime to be a four dimensional smooth manifold $\mathcal{M}$  with metric $g_{\mu\nu}$ of signature $(-,+,+,+)$.  Furthermore, we will take $\mathcal{M}$, or at least a portion of it $M\subseteq\mathcal{M}$, to be globally hyperbolic and such that it may be foliated as $M\approx\mathbb{R}\times\Sigma$, with $\Sigma$ a spatial hypersurface. We will use greek letters to denote abstract or actual component spacetime indices running from 0 to 3. Similarly, lowercase latin letters from the beginning of the alphabet will correspond to abstract or component spatial indices on $\Sigma$ running from 1 to 3. For example, the induced spatial metric on each $\Sigma$ [of signature $(+,+,+)$] is denoted as $q_{ab}$, so if $\{x^\mu\}_{\mu=0,1,2,3}$ are arbitrary coordinates on $\mathcal{M}$ and $\{y^a\}_{a=1,2,3}$ are arbitrary coordinates on $\Sigma$, the vectors (not to be confused with tetrads here):
\[
\tilde{e}^\mu_a:=\frac{\partial x^\mu}{\partial y^a}\,,
\]
form a basis on the tangent space $T_p\Sigma$ at a point $p$ and
\[
q_{ab}=\tilde{e}^\mu_a\,\tilde{e}^\nu_b\,g_{\mu\nu}\,.
\]
The extension of $q_{ab}$ to all $T_p\mathcal{M}$ is then $q_{\mu\nu}=g_{\mu\nu}+n_\mu n_\nu$, with $n^\mu$ the unit time-like normal to $\Sigma$.

Isolated horizons are modeled by special types of null hypersurfaces on $\mathcal{M}$, so we start by reviewing the basic definitions and geometric structures associated with them.  A three dimensional hypersurface $\Delta\subset\mathcal{M}$ is said to be \emph{null} if the induced metric $h:=\iota^*g$ is degenerate and hence necessarily of signature $(0,+,+)$.
Here $\iota:\Delta\hookrightarrow\mathcal{M}$ is the inclusion map so $h$ is the pullback of $g$.
Following previous conventions, we will denote spacetime covariant (or spatial) forms pulled back to $\Delta$ using indices with arrows under them. For consistency in tensorial expressions then --although somewhat cumbersome-- we may also use arrows under indices for elements of tensor products of $T\Delta$. Therefore, analogously to the spatial case, if $\{z^{\pbi{\mu}}\}_{\pbi{\mu}=0,1,2}$ are arbitrary coordinates on $\Delta$ then the vectors:
\[
\tilde{e}^\mu_{\pbi{\mu}}:=\frac{\partial x^\mu}{\partial z^{\pbi{\mu}}}
\]
constitute a basis on the tangent space $T_p\Delta$ and the induced metric is
\[
h_\pbi{\mu\nu}:=h_{\pbi{\mu}\pbi{\nu}}:=g_{\pbi{\mu}\pbi{\nu}}=\tilde{e}^\mu_{\pbi{\mu}}\,\tilde{e}^\nu_{\pbi{\nu}}\,g_{\mu\nu}\,.
\]
Degeneracy of the metric $h_\pbi{\mu\nu}$ is of course equivalent to the requirement that any vector $\ell^\mu$ normal to $\Delta$ (i.e. orthogonal to all vectors in $T_p\Delta$) is null: $\ell_\mu\ell^\mu=0$, and it follows that a fundamental property of null hypersurfaces is that they are ruled by null geodesics. Any null normal vector field $\ell^\mu$ on $\Delta$ satisfies the geodesic equation
\[
\ell^\mu\nabla_\mu\ell^\nu=\kappa_{(\ell)}\ell^\nu\,,
\]
with $\kappa_{(\ell)}$ the \emph{acceleration} or \emph{non-affinity coefficient} of $\ell^\mu$, which corresponds to surface gravity when one specializes to isolated horizons.
The geodesics or more precisely the one dimensional images on $\Delta$ of the integral curves of $\ell^\mu$ are unique, but $\ell^\mu$ is not. A reparameterization of the null geodesic generators defines a different tangent velocity vector field, so any rescaling $\ell^\mu\to\tilde{\ell}^\mu:=\xi\ell^\mu$ for an arbitrary smooth function $\xi:\Delta\to\mathbb{R}$ results in a valid normal.
Under such rescalings $\kappa_{(\ell)}$ transforms as
\[
\kappa_{(\ell)}\to \kappa_{(\tilde{\ell})}=\xi\kappa_{(\ell)}+\ell^\mu\nabla_\mu\xi\,.
\]
Unlike the time-like or space-like case, there is no canonical way to fix the normalization of $\ell^\mu$ to resolve this ambiguity.

\begin{figure}    
     \begin{center}
      \includegraphics[width=8cm]{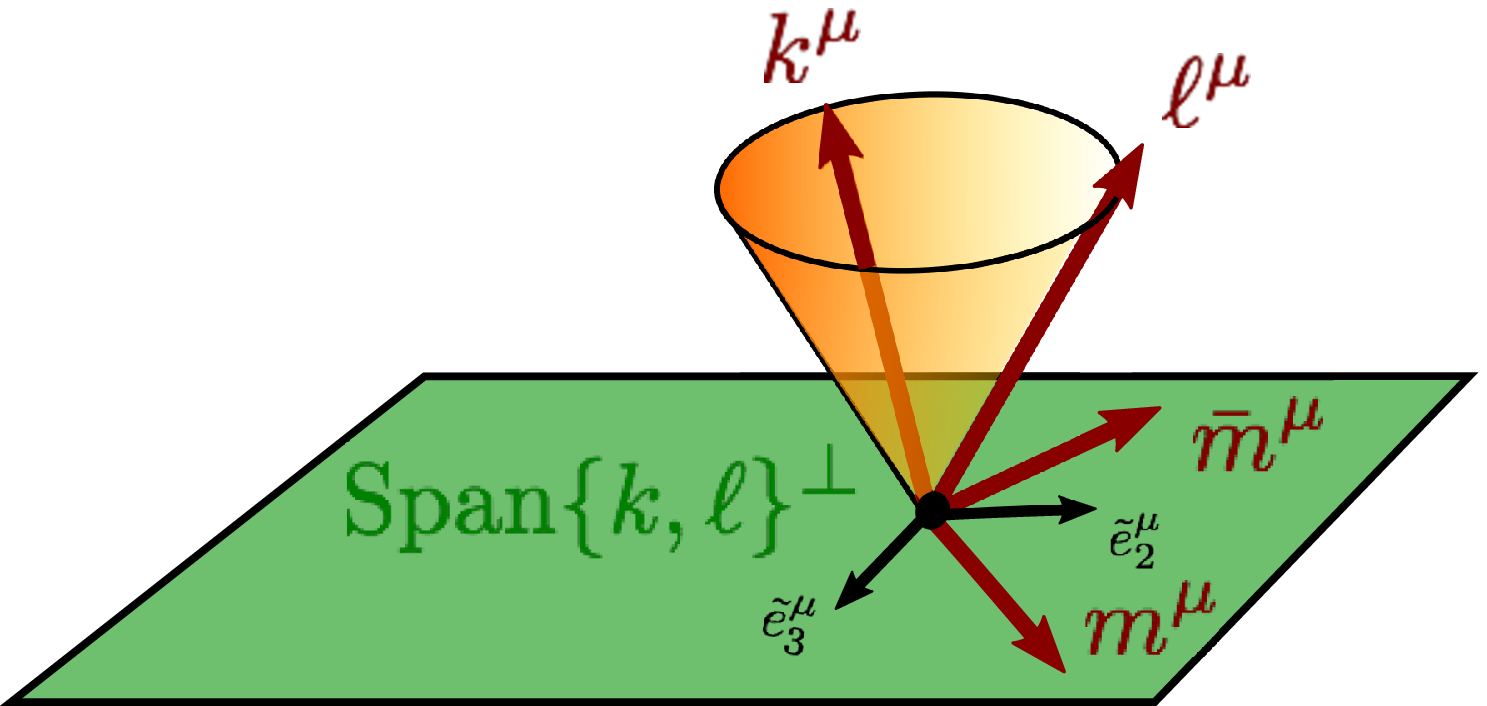}
      \caption{Construction of a Newman-Penrose tetrad $(k^\mu,\ell^\mu,m^\mu,\bar{m}^\mu)$ adapted to null hypersurface $\Delta$.} \label{fNPbasis}
      \end{center}
\end{figure}

For null hypersurfaces, it is convenient to use the null normal $\ell^\mu$  at each point $p\in\Delta$ to construct or fix a \emph{Newman-Penrose null basis}  $(k,\ell,m,\bar{m})$ on $T_p\mathcal{M}$, such that \cite{NewmanPenrose}
\begin{equation} \label{nullProducts}
\begin{array}{ccccc}
k\cdot k=0, &  m\cdot m=0, &  \bar{m}\cdot\bar{m}=0,  & k\cdot \ell=-1, & m\cdot\bar{m}=1, \\
m\cdot\ell=0, &  \bar{m}\cdot\ell=0, & m\cdot k=0, &  \bar{m}\cdot k=0\,. & 
\end{array}
 \end{equation}
 This is done by choosing a  null direction $k^\mu$ transverse to $T_p\Delta$ on the light cone at each $T_p\mathcal{M}$, normalized such that $k_\mu\ell^\mu=-1$ (Figure \ref{fNPbasis}). The orthogonal complement $\text{Span}\{k,\ell\}^\bot$ of the plane spanned by $k^\mu$ and $\ell^\mu$ is a two dimensional spatial subspace of $T_p\Delta$, so for any orthonormal basis $\tilde{e}_2$, $\tilde{e}_3$  on it, one may define the complex vector $m^\mu$ and its conjugate as
\[
m^\mu:=\frac{\sqrt{2}}{2}\left(\tilde{e}_2^\mu+i\tilde{e}_3^\mu\right)\,, \qquad
\bar{m}^\mu:=\frac{\sqrt{2}}{2}\left(\tilde{e}_2^\mu-i\tilde{e}_3^\mu\right)\,.
\]
We will call a basis constructed in this way a null basis \emph{adapted to} $\Delta$.

A null basis adapted to $\Delta$ is everything but unique. Apart from the choice of null normal, there are infinitely many possibilities for the choice of transverse null direction $k^\mu$ and a corresponding  ambiguity in the choice of $m^\mu$ defining the orthogonal spatial plane $\text{Span}\{m,\bar{m}\}=\text{Span}\{k,\ell\}^\bot$, but any two adapted null bases are related by a restricted Lorentz gauge transformation, i.e.  a Lorentz rotation preserving the direction of $\ell^\mu$. It then follows that the \emph{cross sectional area two-form} 
\begin{equation}  \label{area2form}
\leftidx{^2}{\epsilon}{}:=im\wedge\bar{m}\,,
\end{equation}
is invariantly defined, as one can easily verify by transforming the null tetrad via  Lorentz rotations which leave the direction of $\ell$ unchanged (transformations of type I, II and III as defined in the next section).
Similarly, the (two dimensional spatial) \emph{cross sectional metric}  $m\otimes\bar{m}+\bar{m}\otimes m$ is invariantly defined on $T_p\Delta$ and equal to the pullback metric:
\begin{equation}  \label{inducedmetric}
h_\pbi{\mu\nu}=2\,m_{(\pbi{\mu}}\bar{m}_{\pbi{\nu})}\,.
\end{equation}
The pullback metric $h_{\pbi{\mu\nu}}$ may thus be extended to all of $T_p\mathcal{M}$ as
\[
h_{\mu\nu}=2\,m_{(\mu}\bar{m}_{\nu)}=g_{\mu\nu}+2\ell_{(\mu}k_{\nu)}\,.
\]
So the induced metric on $\Delta$ also matches this cross sectional metric `transverse' to the three dimensional congruence of null generators of $\Delta$ and which only acts on the  $\text{Span}\{k,\ell\}^\bot$ part of tangent vectors. We emphasize here that while the area two-form $\leftidx{^2}{\epsilon}{}$ and metric $h_\pbi{\mu\nu}$ are invariantly defined, their corresponding expressions on the right hand side of (\ref{area2form}) and (\ref{inducedmetric}) are only valid if one uses a null tetrad adapted to $\Delta$.

The transversal direction $k^\mu$ allows one to define a projector to $T_p\Delta$
\[
\Pi^\mu_{\;\;\nu}:=\delta^\mu_{\;\;\nu}+k^\mu\ell_\nu\,,
\]
and a projector to $\text{Span}\{k,\ell\}^\bot$
\[
h^\mu_{\;\;\nu}:=\delta ^\mu_{\;\;\nu}+\ell^\mu k_\nu+k^\mu\ell_\nu\,.
\]

The second fundamental form of $\Delta$ with respect to $\ell^\mu$ --the analogous of extrinsic curvature two-form $K_{\mu\nu}$ for spatial (or temporal) hypersurfaces-- is defined as
\begin{equation} \label{ThetaDef}
\Theta_{\mu\nu}:=\Pi^\sigma_{\;\;\mu}\Pi^\rho_{\;\;\nu}\nabla_\sigma\ell_\rho=h^\sigma_{\;\;\mu}h^\rho_{\;\;\nu}\nabla_\sigma\ell_\rho=\nabla_\mu\ell_\nu-\omega_\mu\ell_\nu+\ell_\mu k^\sigma\nabla_\sigma\ell_\nu\,,
\end{equation}
with
\begin{equation}  \label{omegaDef}
\omega_\mu:=-k^\sigma\nabla_\mu\ell_\sigma-\ell_\mu k_\sigma k^\rho\nabla_\rho\ell^\sigma\,,
\end{equation}
defined as the \emph{rotation 1-form} in the context of black hole horizons. Contracting this last definition with $\ell^\mu$, it follows that 
\[
\kappa_{(\ell)}=\omega_\mu\ell^\mu.
\]  

The second fundamental form $\Theta_{\mu\nu}$ encodes the `kinematics' of the geodesic congruence of null generators of $\Delta$ with velocities $\ell^\mu$. Its trace $\theta_{(\ell)}:=g^{\mu\nu}\Theta_{\mu\nu}$ defining the \emph{expansion} of the congruence and its symmetric trace-free part the \emph{shear} (the \emph{twist} or anti-symmetric part being zero as a consequence of the congruence being hypersurface orthogonal).

Strictly, $\ell^\mu$ is only defined on $\Delta$ so to make sense of $\nabla_\mu\ell_\nu$ and define $\Theta_{\mu\nu}$ one requires an extension of $\ell^\mu$ to a neighborhood of $\Delta$. In particular to facilitate comparison with 3+1 structures, one can (and we will) extend $\ell^\mu$ and all geometric structures introduced so far by foliating a neighborhood of $\Delta$ by null hypersurfaces. Such geometric structures on $\Delta$ are independent of the extension. To express relations amongst them which may only hold on $\Delta$, one uses the standard notation $\WIHeq$ for equalities valid only on $\Delta$.


\subsection{Non-expanding horizons}

To model equilibrium horizons resulting from gravitational collapse, one first restricts the topology of $\Delta$ and incorporates the notion that the null geodesic generators should be non-expanding.

A null hypersurface $\Delta\subset\mathcal{M}$ is called a \emph{non-expanding horizon} (NEH) if
\begin{enumerate}[(i)]
\item $\Delta$ is diffeomorphic to the product $S^2\times\mathbb{R}$ with the fibers of the canonical projection $P:S^2\times\mathbb{R}\to S^2$ corresponding to the null generators.
\item The expansion $\theta_{(\ell)}$ of any null normal $\ell^\mu$ to $\Delta$ vanishes.
\item On $\Delta$, Einstein's equations hold and the stress-energy tensor $T_{\mu\nu}$ of matter satisfies the \emph{null dominant energy condition}, i.e. $-T^\mu_{\;\;\nu}\,\ell^\nu$ is causal and future-directed.
\end{enumerate}

Taken together, these conditions imply  that  on $\Delta$ not only the expansion but the complete second fundamental form vanishes $\Theta_{\mu\nu}=0$ \cite{AFK,GJreview}. So from (\ref{ThetaDef}) and $\ell_\pbi{\mu}=0$, it follows that on a NEH
\begin{equation} \label{rotationForm}
\nabla_{\pbi{\mu}}\ell^\nu\WIHeq\omega_\pbi{\mu}\ell^\nu\,.
\end{equation}
This equation first shows that $\omega_\pbi{\mu}$ --the pullback to $T_p^*\Delta$ of the rotation one-form-- is `almost intrisic' to 
$\Delta$ in the sense that it is independent of any transverse direction $k^\mu$.
The form $\omega_\pbi{\mu}$ does however depend on the choice of null normal, transforming as 
\begin{equation} \label{omegaT}
\omega_\pbi{\mu}\to\tilde{\omega}_\pbi{\mu}=\omega_\pbi{\mu}+\nabla_\pbi{\mu}\ln\xi\,,
\end{equation}
for rescalings $\ell\to\xi\ell$. 

Equation (\ref{rotationForm}) also implies \cite{AFK} that the induced metric on $\Delta$ and the transverse area 2-form are Lie dragged along $\ell^\mu$:
\begin{equation} \label{Lh}
\mathcal{L}_{\ell}\,h_{\pbi{\mu\nu}}\WIHeq 0\,  \qquad\text{ and }\qquad  \mathcal{L}_{\ell}\,\leftidx{^2}{\epsilon}{}\WIHeq 0\,.
\end{equation}
This notion of `time independence' of the induced metric is a crucial ingredient in the characterization of black holes in equilibrium but it is not the complete story.
In a general null hypersurface $\Delta$ each choice of transverse direction $k^\mu$ defines -- via the projector and the formula: $\widehat{\nabla}_\mu V^\nu:=\Pi^\alpha_{\;\;\mu}\Pi^\nu_{\;\;\beta}\nabla_\alpha V^\beta$ -- a (torsion-free) induced connection $\widehat{\nabla}_\pbi{\mu}$ compatible with the induced metric: $\widehat{\nabla}_\pbi{\mu}h_{\pbi{\nu\rho}}=0$. Nevertheless, on a NEH this connection is unique and `intrinsic' to $\Delta$, i.e. independent of the choice of transverse direction \cite{AFK}. 
In constrast to the spacelike or timelike case however, on a NEH this connection is not fully determined by the metric. It is hence the pair $\big(h_{\pbi{\mu\nu}},\widehat{\nabla}_\pbi{\mu}\big)$ taken together that defines the \emph{intrinsic geometry} on $\Delta$.

The intrinsic geometry of $\Delta$ determines ${R_{\underset{^{\longleftarrow}}{\mu\nu\rho}}}^\sigma$, the pullback to $\Delta$  of the spacetime Riemann tensor ${R_{\mu\nu\rho}}^\sigma$. Thus the Riemann tensor of a spacetime containing a NEH satisfies important properties whose specific form however will not be directly relevant for our analysis here.
The important point from the perspective of a variational principle is that corresponding configuration spaces consisting of spacetimes $(\mathcal{M},g_{\mu\nu})$ admitting a NEH are still infinite dimensional. Nevertheless,  spacetimes that contain a NEH are clearly a special subset of the configuration space of General Relativity\footnote{Essentially, the condition that no flux of energy may cross a NEH, which may be inferred from its definition \cite{AFK}, effectively reduces by half the number of local (radiative) degrees of freedom \cite{ABF, EngleNouiPerezPranzetti}. This can be understood using a null initial value formulation \cite{Friedrich,Rendall} with `initial' value hypersurface $\mathcal{H}=\Delta\cup\mathcal{N}$ consisting of $\Delta$ and null hypersurface $\mathcal{N}$ transverse to $\Delta$. Radiation from the past domain of dependence of $\mathcal{H}$ can only intersect $\mathcal{N}$, but no transverse radiation may intersect $\Delta$.}.

Apart from the rotation one-form $\omega_\pbi{\mu}$, on a NEH there is another `intrinsic' one-form $V_\pbi{\mu}$ determined by the connection $\widehat{\nabla}_\pbi{\mu}$ that plays an important role in the analysis of isolated horizons in a first order fomulation. Using an adapted null basis, the \emph{transverse connection potential} may be defined as:
\begin{equation} \label{tV}
V_\pbi{\mu}:=\bar{m}_\nu\widehat{\nabla}_\pbi{\mu}m^\nu\,.
\end{equation}
For restricted Lorentz transformations of the null basis, (\ref{tV}) indeed remains invariantly defined except for $U(1)$-rotations (transformations of type II as defined in the next section) for which it transforms as an $U(1)$-connection. Analogously to the area two-form $\leftidx{^2}{\epsilon}{}$ and metric $h_\pbi{\mu\nu}$ and corresponding expressions  (\ref{area2form}) and (\ref{inducedmetric}), the potential $V_\pbi{\mu}$ determines an invariantly defined $U(1)$-connection independent of any null tetrad\footnote{Just as one can regard $\omega_\pbi{\mu}$ as an $\mathbb{R}$-connection potential on the line bundle $T\Delta^\perp$ over $\Delta$ whose fibers are the 1-dimensional null normals to $\Delta$ \cite{AFK}, $V_\pbi{\mu}$ is the potential for an $U(1)$-connection on the `transverse' bundle $T\Delta^\top$ over $\Delta$. 
The 2-dimensional fibers of $T\Delta^\top$ consist of equivalence classes $[X]$ of vectors in $T_p\Delta$, where $X,Y\in T_p\Delta$ are defined to be equivalent if they differ by a multiple of $\ell$:  $X\sim Y$ iff $X-Y\propto\ell$. These classes define non degenerate directions within $\Delta$ `transverse' to the null congruence of its generators (see e.g. Chap. 9 in \cite{WaldGRbook}).
$\omega_\pbi{\mu}$ transforms as in (\ref{omegaT}) under $\mathbb{R}$-rescalings of the null normal: $\ell\to\xi\ell$, whereas $V_\pbi{\mu}$ transforms as $V\to V+i\md\theta$ under $U(1)$-rotations of (the equivalence class of) $m$: $m\to e^{i\theta}m$. 
}.
Expression on the right hand side of (\ref{tV}) is valid only if one uses a null tetrad adapted to $\Delta$.


\subsection{Weakly (and strongly) isolated horizons}

It turns out that to derive the zeroth and first laws of black hole mechanics, one needs to further privilege an equivalence class of normal vector fields on $\Delta$.

A \emph{weakly isolated horizon} $(\Delta,[\ell])$ is a non-expanding horizon where an equivalence class of null normals $[\ell]$ satisfying  
\begin{equation} \label{WIHcondition}
\mathcal{L}_{{\ell}}\,\omega_\pbi{\mu}\WIHeq 0  \qquad \text{ for all } \; \ell^\mu\in[\ell]
\end{equation}
has been singled out. Two normals $\tilde{\ell}\sim\ell$ belong to the same equivalence class iff $\tilde{\ell}^\mu=c\ell^\mu$ for some constant $c>0$ on $\Delta$.  Notice that the choice of equivalence class also singles out a unique rotation 1-form $\omega_\pbi{\mu}$, as transformation (\ref{omegaT}) shows. 

Condition (\ref{WIHcondition}) is equivalent to requiring not only the pullback metric $h_\pbi{\mu\nu}$ but also  the `extrinsic curvature' or \emph{Weingarten  map} ${\mathcal{K}_\mu}^{\nu}:=\widehat{\nabla}_\mu\ell^\nu$ to be time-independent  on $\Delta$.
Indeed, using (\ref{ThetaDef}) and the fact that $\Theta_{\mu\nu}=0$ on a NEH, one has\footnote{On a null hypersurface 
${\mathcal{K}_\mu}^\nu\neq{\Theta_\mu}^\nu$. The Weingarten map and the second fundamental form are instead related by 
${\mathcal{K}_\mu}^\nu={\Theta_\mu}^\nu+\omega_\mu\ell^\nu$.} 
\[
{\mathcal{K}_\mu}^\nu:=\widehat{\nabla}_\mu\ell^\nu:=\Pi^\alpha_{\;\;\mu}\Pi^\nu_{\;\;\beta}\nabla_\alpha\ell^\beta
=\Pi^\alpha_{\;\;\mu}\nabla_\alpha\ell^\nu=\nabla_\mu\ell^\nu+\ell_\mu k^\alpha\nabla_\alpha\ell^\nu\WIHeq\omega_\mu\ell^\nu\,,
\] 
so taking the Lie derivative along $\ell$ and imposing (\ref{WIHcondition}):
\[
\mathcal{L}_{{\ell}}\,{\mathcal{K}_\pbi{\mu}}^\nu\WIHeq\ell^\nu\mathcal{L}_{{\ell}}\,\omega_\pbi{\mu}\WIHeq 0.
\]

This symmetry condition or time-independence of a part of the induced connection
 is sufficient to ensure that the surface gravity $\kappa_{(\ell)}$ is constant on a WIH, that constitutes the zeroth law of black hole mechanics.
As shown in \cite{AFK}, this follows by using a relation valid for NEHs:
\begin{equation}  \label{domega}
\underleftarrow{\md\omega}\WIHeq 2\,\rm{Im}[\Psi_2]\,\underleftarrow{\leftidx{^2}{\epsilon}}\,,
\end{equation}
where 
$\Psi_2:\WIHeq C_{\mu\nu\rho\sigma}\ell^{\mu}m^{\nu}\bar{m}^{\rho}k^{\sigma}$ is the (gauge invariant) Weyl tensor component. Since $\ell\,\cdot\,\leftidx{^2}{\epsilon}
\WIHeq 0$ then
\begin{equation}
\md \kappa_{(\ell)}=\md (\ell\cdot\omega )\WIHeq  \md (\ell\cdot\omega )+\ell\cdot\md\omega =\mathcal{L}_{\ell}\,\omega\WIHeq 0\,, 
\end{equation}
where to avoid cluttering notation, all forms are assumed to be pulled back to $\Delta$.

Additionally, as it was also shown in \cite{AFK}, a first law can be derived for this class of horizons with preferred equivalence class $[\ell]$  satisfying (\ref{WIHcondition}). 

Every NEH can be made into a WIH by  appropriate (albeit infinitely many different) choices of equivalence classes $[\ell]$.  We will hence work mainly within the class of weakly isolated horizons in this paper.

 A more restricted proper subclass of WIHs is that of strongly isolated horizons (or simply isolated horizons). This latter class models --perhaps more accurately-- the notion of a horizon in equilibrium by demanding the whole intrinsic geometry of $\Delta$ to be independent of time.

More precisely, we will call \emph{strongly isolated horizon} (SIH) a WIH $(\Delta,[\ell])$ for which the induced connection $\widehat{\nabla}$ further satisfies:
\begin{equation} \label{SIHcond}
[\mathcal{L}_{{\ell}}\,,\widehat{\nabla}]\WIHeq0\,.
\end{equation}
That this condition really captures the notion of time independence of the connection $\widehat{\nabla}$ can be understood by analyzing the free data and constraints on WIHs \cite{ABLgeometry}.

Finally, we recall that weakly (and strongly) isolated horizons may be classified according to their symmetries or the group of diffeomorphisms on $\Delta$ which preserve the horizon structure, i.e. the equivalence class $[\ell]$ and the geometry $(h_\pbi{\mu\nu},\omega_\pbi{\mu})$ (or the full connection $\widehat{\nabla}_\pbi{\mu}$ for SIHs). 
Specifically, an infinitesimal generator $W^\mu\in T\Delta$ of such a symmetry diffeomorphism of a WIH satisfies:
\[
\mathcal{L}_W\ell^\mu\WIHeq \delta c\, \ell^\mu, \quad\text{with }  \delta c \text{ constant on } \Delta, 
 \qquad \mathcal{L}_Wh_\pbi{\mu\nu}\WIHeq 0 
 \quad \text{ and } \quad \mathcal{L}_W\omega_\pbi{\mu}\WIHeq 0\,.
\]
For the non extremal case $\kappa_{(\ell)}\neq 0$, the Lie algebra of infinitesimal symmetry generators is finite dimensional and there are only three possibilities \cite{ABL}: 
\begin{enumerate}[Type I)]
\item Geometry is spherically symmetric: Lie algebra is four dimensional and equals the semi direct sum of constant rescalings  $W^\mu=c\ell^\mu$ and rotation generators $\mathfrak{so}(3)$.
\item Geometry is axi-symmetric: Lie algebra is two dimensional and equals the semi direct sum of constant rescalings  $W^\mu=c\ell^\mu$ and rotation generators  $\mathfrak{so}(2)$ around an axis.
\item Distorted or generic: Lie algebra consists only of the one dimensional constant rescalings  $W^\mu=c\ell^\mu$.
\end{enumerate}
For the extremal case $\kappa_{(\ell)}=0$, there are the same three possibilities, except that the one dimensional Lie algebra of constant rescalings  [isomorphic to $(\mathbb{R},+)$] is replaced by the infinite dimensional Lie algebra of rescalings $W^\mu=f\ell^\mu$, such that $\mathcal{L}_\ell f=$ constant. 


\subsection{Free data on a NEH, WIH and SIH}

An analysis of the free data and constraints on NEHs and IHs was first undertaken from an intrinsic perspective in \cite{ABLgeometry}  and more extensively in \cite{GJreview} from a 3+1 perspective. The freely specifiable geometric data on a NEH can be given in terms of fields living in any spatial two-sphere cross section of $\Delta$. This is most appropriate for a 3+1 decomposition and/or an initial value problem formulation. The two dimensional fields are Lie-dragged along $\ell$ according to certain \emph{constraint equations} in order to reconstruct the full geometry of $\Delta$.
 
Let a spatial Cauchy hypersurface $\Sigma_{t_0}$ intersect $\Delta$ in the sphere $\mathcal{S}_{t_0}$. This singles out a transverse direction $k^\mu$ as the unique null normal to $\mathcal{S}_{t_0}$ such that $k\cdot\ell=-1$. 
By Lie dragging $\mathcal{S}_{t_0}$ along $\ell^\mu$ one may construct a foliation and extend $k^\mu$ to all (or at least a portion of) $\Delta$\footnote{We do not worry about completeness of $\ell$ here.}.
The pullback  of the intrinsic metric $h_\pbi{\mu\nu}$ to the sphere $\mathcal{S}_{t_0}$ --which we denote\footnote{Consistent with the index notation for spatial tensors on $\Sigma_{t_0}$, we use latin letters from the beginning of the alphabet for indices of forms on $\Delta$ pulled back to the two-sphere $\mathcal{S}_{t_0}$, even though here they are really two dimensional indices rather than three dimensional.}  as $h_{ab}$-- is the two dimensional spatial cross sectional metric and it is freely specifiable. The metric $h_\pbi{\mu\nu}$ may be reconstructed from $h_{ab}$ using the constraint equation (\ref{Lh}).

On the other hand, the metric $h_{ab}$ determines a connection $\leftidx{^2}{D}{_a}$ on the sphere which coincides with $\widehat{\nabla}_\pbi{\mu}$ for vectors on $T_p\mathcal{S}_{t_0}$. For arbitrary  vectors on $T_p\Delta$, the action of the connection $\widehat{\nabla}_\pbi{\mu}$ in transverse spatial-null directions  may be reconstructed from $\leftidx{^2}{D}{_a}$ and $\nabla_\pbi{\mu} k_\pbi{\nu}$. The action of the connection along $\ell^\mu$ is determined by $\omega_\pbi{\mu}$.
In short, $\widehat{\nabla}_\pbi{\mu}$ may be reconstructed from the data $(\leftidx{^2}{D}{_a}, \omega_\pbi{\mu}, \nabla_\pbi{\mu} k_\pbi{\nu})$. 

In turn, $\omega_\pbi{\mu}$ and  $\nabla_\pbi{\mu} k_\pbi{\nu}$ may be written in terms of two dimensional fields plus $\kappa_{(\ell)}$ and $k^\mu$:
\begin{equation} \label{HajicekForm}
\omega_\mu=\Omega_\mu-\kappa_{(\ell)}k_\mu\,,
\end{equation}
\[
\nabla_\mu k_\nu=\Xi_{\mu\nu}-k_\mu\Omega_\nu-\ell_\nu k^\sigma\nabla_\sigma k_\mu-\omega_\mu k_\nu\,,
\]
where $\Omega_\mu:=h^\nu_{\;\mu}\omega_\nu$ is the projection  of the rotation 1-form to the sphere and which also coincides\footnote{The pullbacks to the sphere $\ell_a$ and $k_a$ vanish and the `pullback' $h^\mu_{\;\;a}$ of the projector $h^\mu_{\;\;\nu}$ is the identity.} with its pullback $\omega_a$ on $T_p\mathcal{S}_{t_0}$: $\omega_a=\Omega_a$. 
This is called the \emph{H\'a\'{\j}i\v{c}ek 1-form} \cite{Hajicek}.
$\Xi_{\mu\nu}:=h^\sigma_{\;\;\mu}h^\rho_{\;\;\nu}\nabla_\sigma k_\rho$ is the \emph{transversal deformation rate}, the analog for $k^\mu$ of the second fundamental form (\ref{ThetaDef}).
$\Omega_a$ and $\Xi_{ab}=\nabla_ak_b$ are completely unrestricted and $\Omega_\mu$ and $\Xi_{\mu\nu}$ may be determined from them  on all of $\Delta$ using certain constraint equations for $\mathcal{L}_\ell\,\Omega_\mu$ and $\mathcal{L}_\ell\,\Xi_{\mu\nu}$ whose exact form is given in \cite{GJreview, ABLgeometry}.

In summary, the intrinsic geometry $(h_\pbi{\mu\nu},\widehat{\nabla}_\pbi{\mu})$ on a NEH is completely determined by two dimensional fields on a cross sectional sphere $\mathcal{S}_{t_0}$ and the surface gravity:
\[
(h_{ab}, \Omega_a, \Xi_{ab}, \kappa_{(\ell)})\,,
\]
all of which may be arbitrarily specified.

If we now specialize to WIHs, the arbitrariness in the choice of surface gravity is restricted by the zeroth law: $\kappa_{(\ell)}(p)=\kappa_0$ for all $p\in\Delta$, so that it becomes a global parameter:  
\begin{equation}    \label{WIHdata}
(h_{ab}, \Omega_a, \Xi_{ab}, \kappa_{0})\,.
\end{equation}
Variations of the WIH geometry result in variations of these free parameters:
\begin{equation}    \label{WIHgVar}
(\delta h_{ab}, \delta \Omega_a, \delta\Xi_{ab}, \delta \kappa_{0})\,.
\end{equation}

The full time independence of the connection on a SIH (\ref{SIHcond}) translates into the additional condition \cite{ABLgeometry, GJreview}
\[
\mathcal{L}_\ell\,\nabla_\pbi{\mu}k_\pbi{\nu}=0\,,
\]
which implies in particular $\mathcal{L}_\ell\,\Xi_{ab}=0$.
The constraint equation for $\mathcal{L}_\ell\,\Xi_{\mu\nu}$ now becomes an actual constraint on the geometry (\ref{WIHdata})
from which, in the non-extremal case $\kappa_{(\ell)}\neq 0$, $\Xi_{ab}$ can be completely specified in terms of the other fields
(for details, including the extremal case, see \cite{GJreview}).  
So the free data on a SIH are:
\[
(h_{ab}, \Omega_a, \kappa_{0})\,.
\]

Regarding the arbitrariness of the cross sectional metric $h_{ab}$ and its constancy along $\ell$, one can now see the \emph{cross sectional area} on a NEH:
\[
a_\Delta:=\int_{\mathcal{S}_{t_0}}\md^2z \, \sqrt{\det h_{ab}}\;=\int_{\mathcal{S}_{t_0}}\leftidx{^2}{\epsilon}{}\,,
\]
is independent of the cross sectional sphere $\mathcal{S}_{t_0}$ (and therefore independent of a foliation of $\Delta$).
Indeed, any other cross section $\mathcal{S}_{t_1}$ can be obtained by Lie dragging $\mathcal{S}_{t_0}$ in the direction of $\ell^\mu$, i.e. along the integral curves $\gamma^\mu(\lambda,z)$ of some vector field $\xi\ell^\mu=\frac{\md\gamma^\mu}{\md\lambda}$ where $z$ coordinatizes points on the sphere [here $\xi(\lambda,z)$ is some generally non constant function on $\Delta$]. Now the result follows from the constancy of the transverse cross sectional metric or the area two-form (\ref{Lh}) and orthogonality $\leftidx{^2}{\epsilon}{}\cdot\ell=0$:
\[
\mathcal{L}_{\xi\ell}\,\leftidx{^2}{\epsilon}{}=\xi\,\mathcal{L}_{\ell}\,\leftidx{^2}{\epsilon}{}\WIHeq 0\,,
\]
for arbitrary function $\xi:\Delta\to\mathbb{R}$.  So obviously, fixing the global parameter $a_\Delta$ partially fixes the freedom in the choice of transverse metric $h_{ab}$ by constraining its determinant, but the other parameters remain free.

Note that for Type I WIHs or SIHs, the only freedom in the specification of the spherically symmetric metric is the aerial radius.    Variations of the cross sectional metric $\delta h_{ab}$ reduce to variations of the cross sectional area $\delta a_\Delta$.

\subsubsection{Restriction to good cuts}  \label{S_goodcuts}

As we shall recall more thoroughly in section \ref{S_decomposition}, there is considerable freedom in the choice of a foliation of $\Delta$ which is in some sense \emph{compatible} with the WIH equivalence class $[\ell]$. As we have already seen, a foliation of $\Delta$ selects a preferred transverse vector field $k^\mu$, so from equation (\ref{HajicekForm}), it is seen that this ambiguity may be linked to the freedom in the choice of the H\'a\'{\j}i\v{c}ek 1-form\footnote{By definition this form depends on the foliation.}. 
In particular, for a fixed WIH geometry $(h_\pbi{\mu\nu},\widehat{\nabla}_\pbi{\mu})$ --and hence a fixed $\omega_\pbi{\mu}$-- a choice of $\Omega_a$ defines a foliation and viceversa.

On a general two-sphere $\mathcal{S}_t$, the one form $\Omega_a$ can always be split into a divergence free part plus an exact form\footnote{This follows from the general Hodge theory on compact manifolds \cite{Nakahara}.}:
\[
\Omega_a=\Omega_a^{\text{div-free}}+\leftidx{^2}{D}{}_af_1\,,
\]
so that $\leftidx{^2}{D}{}_a\Omega^a=\leftidx{^2}{D}{}_a\leftidx{^2}{D}{}^af_1$, for some function $f_1:\mathcal{S}_t\to\mathbb{R}$.
In turn, the divergence free part may be written as $\Omega_a^{\text{div-free}}=\leftidx{^2}{\epsilon}_{ab}\,\leftidx{^2}{D}{^b}f_2$, for another function $f_2:\mathcal{S}_t\to\mathbb{R}$.
Furthermore, using $(\md\omega)_{ab}=(\md\Omega)_{ab}=(\md\Omega^{\text{div-free}})_{ab}$ and relation (\ref{domega}), the function $f_2$ can be related to the Weyl curvature coefficient $\Psi_2$ as:
\[
\leftidx{^2}{D}{}_a\leftidx{^2}{D}{}^af_2=2\,\rm{Im}[\Psi_2]\,.
\]
Hence, the freedom in the choice of $\Omega_a$ is encoded in the two real functions $f_1$ and $f_2$ (or equivalently $\rm{Im}\,\Psi_2$). 

For a given WIH geometry, $\Psi_2$ is fixed, so the H\'a\'{\j}i\v{c}ek 1-form --and with it a foliation of $\Delta$-- is completely determined by specifying its divergence. 
A  foliation may be singled out by requiring $\leftidx{^2}{D}{}_a\Omega^a=0$.
It is called a \emph{foliation into good cuts}. These particular foliations can be considered natural for type I WIHs \cite{ABLgeometry}, but for the general case there is no a priori reason to choose such foliations, e.g. for the Kerr horizon, a foliation into good cuts does not match the usual Kerr-Schild slicing (see \cite{GJreview} and references therein). So for a general analysis or a 3+1 decomposition, one may prefer not to restrict the foliation. We will not generally require a foliation into good cuts in this work.

What the discussion above does make self-evident is that restricting to foliations into good cuts further reduces the free data on a WIH. When varying the WIH geometry, preserving the good cuts condition, implies variations $\delta \Omega_a$ are `cut in half' and replaced by variations of a single function $\delta f_2$ on the sphere, or equivalently $\delta(\rm{Im}[\Psi_2])$\footnote{On a WIH: $\mathcal{L}_\ell\,\rm{Im}[\Psi_2]=0$, so indeed $\rm{Im}[\Psi_2]$ is effectively a function on the sphere too.}.  
In particular, for Type I WIHs  one has $\rm{Im}[\Psi_2]=0$, so restricting to a foliation into good cuts sets $\Omega_a=0$. 
Specializing further to Type I SIHs, the freedom left then reduces to the global parameters:
$(\delta a_\Delta, \delta\kappa_0)$. In other words, there is a single two-parameter family of Type I SIHs with a preferred foliation into good cuts. The freedom in the value of surface gravity may further be fixed by picking a particular element $\ell^\mu$ in the equivalence class $[\ell]$, so with the restriction to good cuts, there is essentially a single Type I SIH geometry on the phase space of general relativity (the one-parameter family of spherically symmetric metrics on the sphere).
This result is key to the viability of the variational principle without any gauge fixing in \cite{ENPprl, EngleNouiPerezPranzetti} for Type I SIHs.


To end this section, we emphasize  --as already pointed out in \cite{EngleNouiPerezPranzetti}-- that all definitions above are purely geometric, not involving, and hence by themselves not restricting any additional `internal' or gauge degrees of freedom which may be used to describe the geometry (as for example those of a first order formulation of gravity in terms of tetrads and connections). However, while the (weakly) isolated horizon boundary conditions on the fields alone do not restrict the gauge, other conditions, like the requirement of a well posed variational principle or Hamiltonian formulation may do so.


\section{First order formulation}  \label{S_FirstOrder}

In order to derive the first law for weakly isolated horizons or to apply (canonical) quantization and perform a statistical quantum mechanics calculation of the celebrated area-entropy formula, one needs to start with a well defined action and variational principle for general relativity which accommodates spacetime solutions admitting WIHs as boundaries.
The specification of such action principle  entails an accurate description of the configuration or phase space considered.
This description includes the set of variables to describe the gravitational field on each spacetime $\mathcal{M}$, the region of integration and its boundaries in the action, along with conditions for the fields on these boundaries.

\begin{figure}    
     \begin{center}
      \includegraphics[width=6cm]{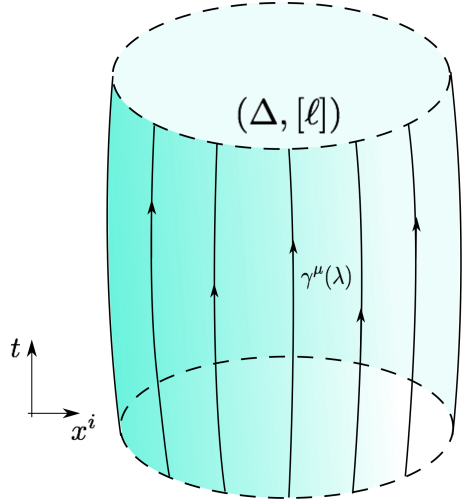}
      \caption{Fixed geometric data on inner boundary $\Delta$.} \label{fWIHgeometry}
      \end{center}
\end{figure}

In geometric terms, the most appropriate configuration space for such action principles consists of asymptotically flat\footnote{We will assume asymptotic flatness for concreteness in the variational principle albeit most of our discussions and conclusions following from horizon boundary conditions are insensitive to this choice.} spacetimes $(\mathcal{M},g_{\mu\nu})$ with a fixed  hypersurface $\Delta$ with generator curves $\gamma^\mu(\lambda)$ whose velocity vector field $\frac{\md\gamma^\mu}{\md\lambda}$ and its constant multiples  define an equivalence class denoted $[\ell]$. Only metrics $g_{\mu\nu}$ which make $(\Delta, [\ell])$ a WIH will be allowed. $\mathcal{M}$ is not a manifold with boundary, but the hypersurface $\Delta$ will act as an inner boundary for the variational principle (Figure \ref{fWIHgeometry}). 

With an eye on quantization, we consider here specific first order formulations of gravity. We focus on the modern description, appropriate as the starting point for loop quantization, where the independent fields are (the components of) the co-tetrad one-form $e^I$ and a real Lorentz connection potential $\A^I_{\;\,J}$. We will also comment on the self-dual formulation as we move along.
 The co-tetrad one-form $e^I=e_{\mu}^I\md x^\mu$  determines the spacetime metric via 
 \begin{equation}
 g_{\mu\nu}=\eta_{IJ}e_\mu^Ie_\nu^J\,,
 \end{equation}
and defines a vector space isomorphism between the tangent space $T_p\mathcal{M}$ at any spacetime point $p$ and a fixed \emph{internal}  Minkowski space with metric $\eta_{IJ}$. Its inverse is the tetrad field $e^\mu_I$. Here and in what follows, latin capital letters from the middle of the alphabet will  denote abstract or actual component indices with respect to some fixed basis of this internal Minkowski space.  
The connection potential one-form $\A^I_{\;\,J}=\A^I_{\mu\,J}\,\md x^\mu$ is $\mathfrak{so}(1,3)$ Lie algebra-valued, so that $\A_\mu^{IJ}=-\A_\mu^{JI}$. The connection defines a covariant derivative acting on internal indices
\begin{equation} \label{connectionPotentialDef}
\leftidx{^\A}{\mathcal{D}}{_\mu}X^I:=\bar{\partial}_\mu X^I+\A^I_{\mu\,J}X^J\,,
\end{equation}
for any other flat and torsion free covariant derivative operator $\bar{\partial}_\mu$.

Following the original treatment for weakly isolated horizons \cite{AFK}, one may fix an \emph{internal} Newman-Penrose null tetrad field $(k^I,l^I,m^I,\bar{m}^I)$ satisfying relations analogous to (\ref{nullProducts}) on the internal Minkowski space at each point on $\Delta$. On $\Delta$ then, the derivative operator $\bar{\partial}_\mu$ in (\ref{connectionPotentialDef}) is chosen so that it annihilates each element of the null basis: $\bar{\partial}_\mu(k^I,l^I,m^I,\bar{m}^I)=0$.
(On the asymptotically flat region of spacetimes $\mathcal{M}$,  boundary conditions imply co-tetrad fields $e_\mu^I$ asymptote to a fixed co-tetrad $\leftidx{^0}{e}{_\mu^I}$ and $\bar{\partial}_\mu$ is chosen to annihilate the latter).

\begin{figure}[h]  
       \begin{center}
       \subfigure[{Adapted Null gauge: The orthonormal tetrad $e^\mu_I$ is restricted so that NP basis is adapted to $\Delta$ and $l^\mu\in[\ell]$.} ]{
      \includegraphics[width=6cm]{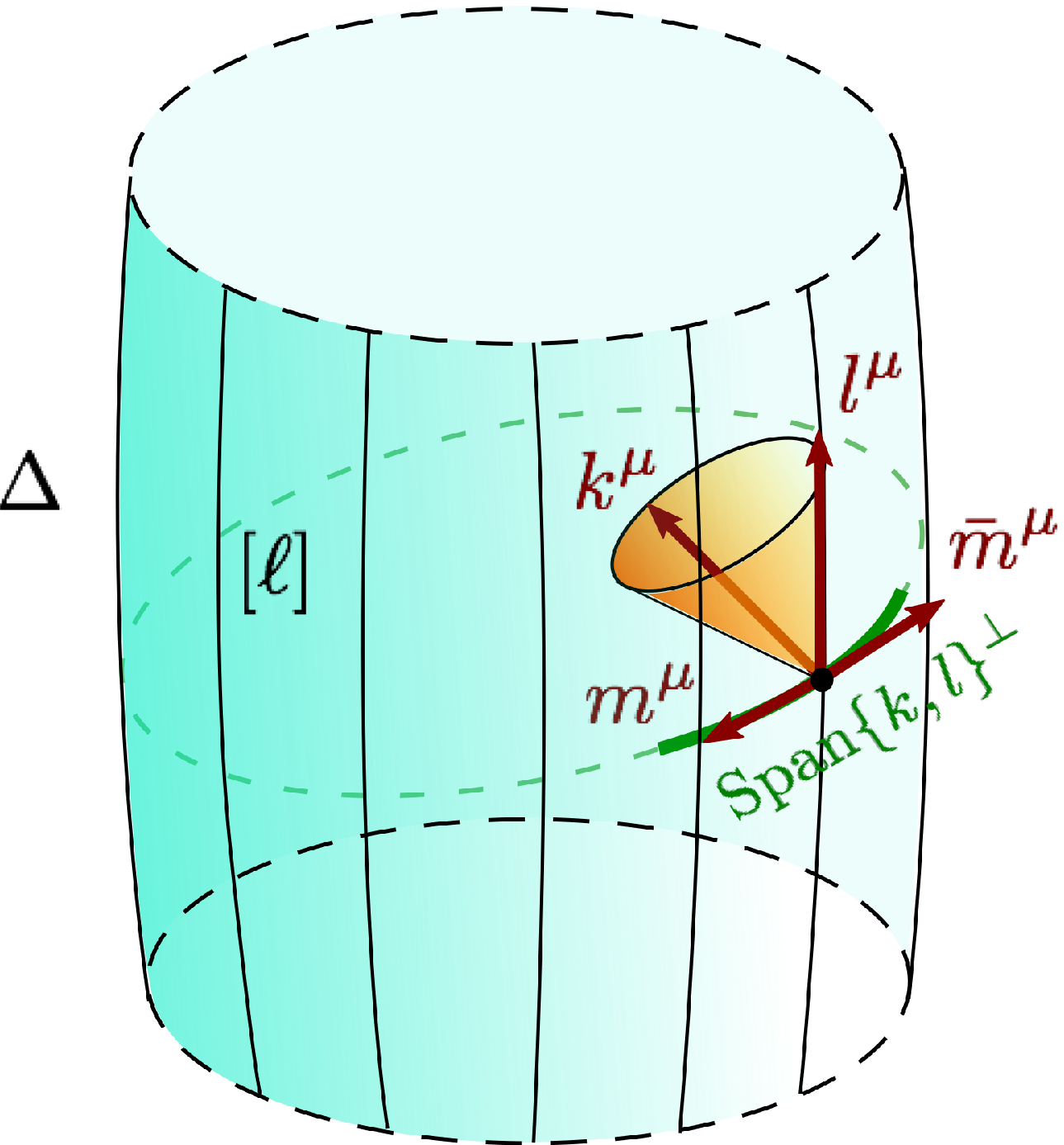}  
           } 
      \hspace{2cm}
    \subfigure[{No gauge fixing: Orthonormal tetrad $e^\mu_I$ is unrestricted so in general $l^\mu\notin[\ell]$ and NP basis need not be adapted to $\Delta$.}]{ 
    \includegraphics[width=7.2cm]{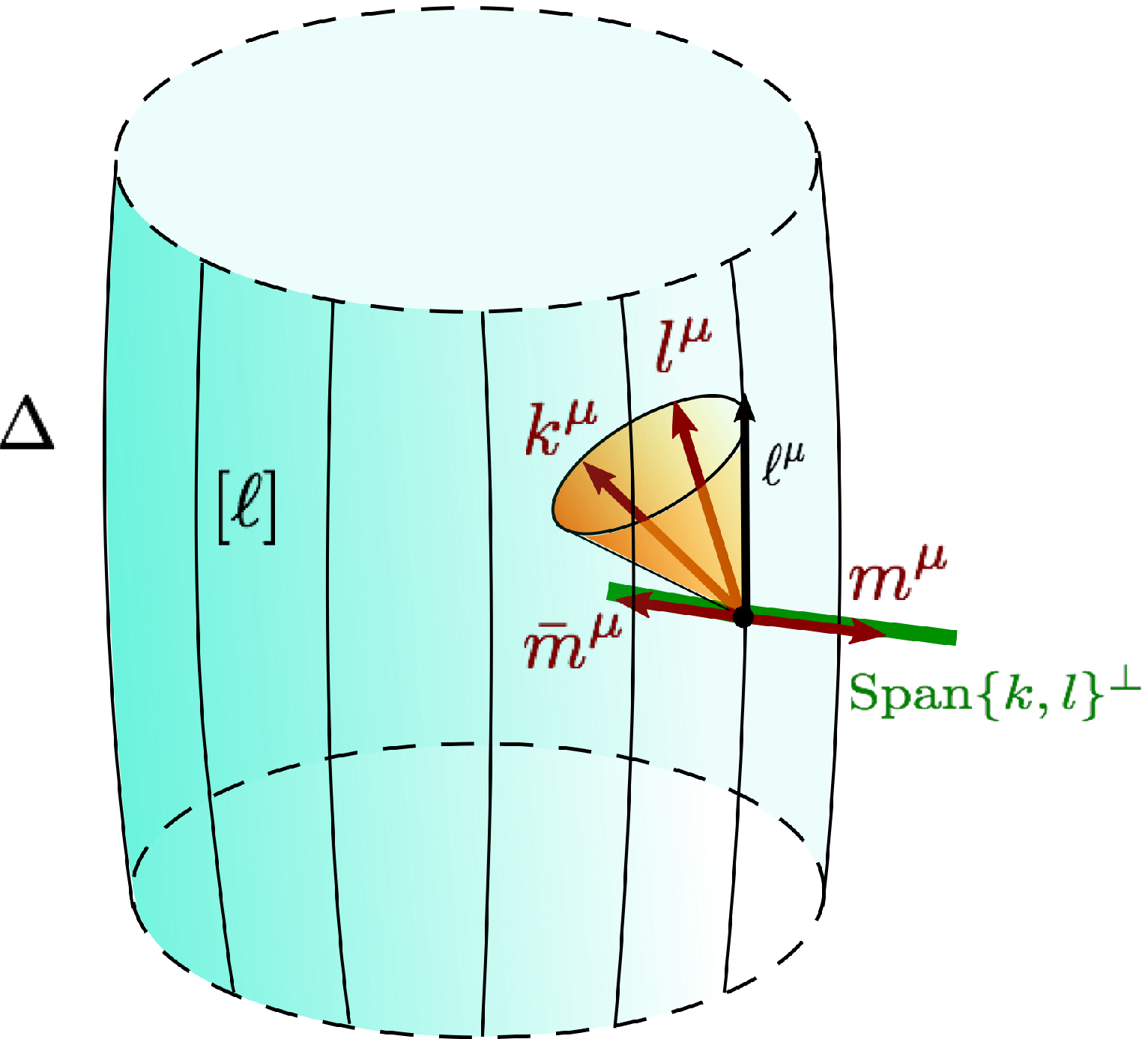}  
    }
    \caption{Spacetime Newman-Penrose null basis $(k^\mu,l^\mu,m^\mu,\bar{m}^\mu)$ defined by isomorphism (\ref{spacetimeNPtetrad}).} \label{fAFKgauge}
    \end{center}
    \end{figure}

By the isomorphism between internal Minkowski spaces and the tangent spaces $T_p\mathcal{M}$ induced by the tetrad, the internal Newman-Penrose null basis also defines a \emph{spacetime} Newman-Penrose null basis via:
\begin{equation} \label{spacetimeNPtetrad}
k^\mu:=e^\mu_Ik^I\,, \quad
l^\mu:=e^\mu_Il^I\,, \quad 
m^\mu:=e^\mu_Im^I\,, \quad
\bar{m}^\mu:=e^\mu_I\bar{m}^I\,.
\end{equation}
A priori however, this null basis may not be linked or adapted to the null surface $\Delta$, i.e. $l^\mu$ --as defined in (\ref{spacetimeNPtetrad})-- need not be normal to $\Delta$ (Figure \ref{fAFKgauge}). To emphasize this point, we have used a different notation for the null vector $l^\mu$ defined this way. For the rest of the paper $\ell^\mu$ will always denote an element of the equivalence class: $\ell^\mu\in[\ell]$ which is invariantly defined up to constant global rescalings.

The original boundary conditions for the fields on $\Delta$ used in \cite{AFK} are:
\begin{enumerate}[(i)]
\item  The pair $(e^I,\A_{IJ})$ is such that $(\Delta,[\ell])$ is a weakly isolated horizon.
\item \emph{Adapted Null gauge}: Tetrads are restricted so that the vector $l^\mu:=e^\mu_I l^I$ always belongs to the equivalence classs $[\ell]$. \label{bcond}
\end{enumerate}
Condition (\ref{bcond}) -- along with the restriction to a fixed internal null basis, so that 
($\delta k^I,\delta l^I,\delta m^I,\delta\bar{m}^I)=0$ -- is a technical restriction necessary to have variations of the 
fields on $\Delta$ under control. 
It implies, as we shall review, a gauge reduction in the first order description of the theory.

\subsection{Gauge symmetries}  \label{S_gauge}

The first order description of the gravitational field through the pair $(e^I,\A_{IJ})$ allows  for the freedom of local Lorentz gauge transformations of the fields:
\[
e_\mu^I\to \tilde{e}_\mu^I:=\Lambda^I_{\;\;J}\,e_\mu^J    \qquad \text{ and } \qquad \A^I_{\mu\,J}\to\tilde{\A}^I_{\mu\,J}:=\Lambda^I_{\;\;K}\,\A^K_{\mu\, L}\,(\Lambda^{-1})^L_{\;\;J}\,+\,\Lambda^I_{\;\;K}\,\partial_\mu(\Lambda^{-1})^K_{\;\;J}\,,
\]
with $\Lambda^I_{\;\;J}\,\in SO(1,3)$.
On $\Delta$ however, boundary condition (\ref{bcond}) or the requirement that $l^\mu:=e^\mu_I l^I\in[\ell]$ for all tetrads, partially reduces this gauge at each point in $\Delta$ to a subgroup which may be expressed as semidirect product $\mathbb{R}^+_{\text{global}}\rtimes ISO(2)$ \cite{BasuChatterjeeGhosh}.

Indeed, given a null tetrad basis $(k, l, m, \bar{m})$  (in the internal or tangent space),  Lorentz transformations ``rotating" the null tetrad at each point $p\in\Delta$ may be conveniently decomposed or expressed as the composition of four types of transformations \cite{Chandrasekhar}:
\begin{enumerate}[(I)]
\item Re-scalings of $k$ and $l$:    
\[
l\to \xi l, \quad k\to\xi^{-1}k, \quad  m\to m, \quad \bar{m}\to\bar{m},  \qquad \text{ for } \xi\in\mathbb{R}\backslash\{0\}.
\]
\item Rotations of $m$ and $\bar{m}$ preserving $k$ and $l$
\[
l\to l, \quad k\to k, \quad  m\to e^{i\theta}m,   \quad  \bar{m}\to e^{-i\theta}\bar{m},   \qquad \text{ for } \theta\in\mathbb{R}.
\]
\item Rotations of $k$ leaving $l$ unchanged
\[
l\to l, \quad k\to k+\bar{a}m+a\bar{m}+a\bar{a} l,  \quad m\to m+a l,  \quad \bar{m}\to \bar{m}+\bar{a} l,
 \qquad \text{ for } a\in\mathbb{C}.
\]
\item Rotations of $l$ leaving $k$ unchanged
\[
l\to l+\bar{b}m+b\bar{m}+b\bar{b}k, \quad k\to k, \quad  m\to m+bk,  \quad  \bar{m}\to\bar{m}+\bar{b}k,
 \qquad \text{ for } b\in\mathbb{C}.
\]
\end{enumerate}
With real $\xi$, $\theta$ and complex numbers $a$, $b$ giving the 6 real parameters for the Lorentz group.

On the other hand, 
while an internal null tetrad $(k^I,l^I,m^I,\bar{m}^I)$ remains fixed,  a gauge transformation acting on the tetrad 
$e^\mu_I\to \tilde{e}^\mu_I:=\Lambda_I^{\;\;J}\,e^\mu_J$ (where $\Lambda_I^{\;\;J}$ denotes the inverse of $\Lambda^I_{\;\;J}$)  
changes or induces a rotation of the spacetime null basis defined by the isomorphism (\ref{spacetimeNPtetrad}):
\[
(k^\mu,l^\mu,m^\mu,\bar{m}^\mu)\; \to \;  (\tilde{k}^\mu,\tilde{l}^\mu,\widetilde{m}^\mu,\bar{\widetilde{m}}^\mu):=
(\tilde{e}^\mu_Ik^I,\tilde{e}^\mu_I l^I,\tilde{e}^\mu_Im^I,\tilde{e}^\mu_I\bar{m}^I)\,.
\]
The requirement that not only $l^\mu:=e^\mu_I l^I$ but also $\tilde{l}^\mu:=\tilde{e}^\mu_I l^I$ belong to the equivalence 
class $[\ell]$ defining a WIH translates to $\tilde{l}^\mu=c\,l^\mu$,  for some constant  $c\in\mathbb{R}^+$,
which restricts Lorentz transformations to those preserving the direction of the null vector $l^\mu$.
That is, only the 4-parameter transformations of type I, II and III are allowed. Transformations of type I generate 
$\mathbb{R}^+$, while type II and III correspond to transformations in the \emph{little group} of null vector $l^\mu$ 
(intuitively, this is the group of rotations in the $k-m$, $k-\bar{m}$ and $m-\bar{m}$ planes leaving $l^\mu$ unchanged) 
which is isomorphic to the inhomogeneous group of rotations in two dimensions $ISO(2)$. The latter transformations are truly 
local, but the requirement for $l^\mu$ to be not only normal to $\Delta$ but to belong to the fixed equivalent class $[\ell]$ 
restricts type I transformations to being only global on $\Delta$.

One may regard this gauge reduction implied by boundary condition (\ref{bcond}) as harmless or inconsequential in deriving 
the classical results. After all, as it has already been emphasized here and in the literature, WIH conditions on a 
hypersurface $\Delta$ are purely geometrical and their consequences for a spacetime geometry should be independent of 
any gauge or gauge-fixed formulation used to describe them. The partial gauge fixing may then be regarded as a `trick' 
that takes advantage of the fact that, on a spacetime containing a WIH \emph{there is} a preferred direction at points 
on $\Delta$, namely the direction along the null generators $\gamma^\mu$. Nevertheless, the same geometric nature of 
WIHs can be used to argue against any gauge fixing. This is of course specially relevant in the quantum theory, where 
-aside from the fact that `$U(1)$ and $SU(2)$ descriptions' render different results \cite{ABCKprl, ENPprl}- a first order 
formulation of gravity is necessary to couple to fermions and the purely geometric conditions for a WIH should not restrict 
local Lorentz invariance. One could (and should) therefore try to relax such condition which `anchors'  $l^\mu:=e^\mu_I l^I$ 
so that it belongs to the equivalence class $[\ell]$. This is a restriction for the tetrad and certainly not a requirement 
for a metric admitting a WIH, so one could in principle allow for more general tetrads for which the null vector $l^\mu$ might 
not even be a normal to $\Delta$. In the following section we shall analyse the necessity of imposing the condition (ii) for 
the formulation of a well posed variational principle.

At first sight, condition (\ref{bcond}) may seem analogous to the boundary condition at infinity which limits tetrads 
to those which asymptote to a fixed tetrad $\leftidx{^0}{e}{^\mu_I}$. Nevertheless, both conditions are very different in nature. 
Restrictions on the tetrad at infinity are solely and directly related to conditions on the metric $g_{\mu\nu}$. Thanks to the 
essential uniqueness of the asymptotic metric --a Minkowski or flat metric-- all points in configuration space asymptote to, 
variations at infinity are well under control. The gauge freedom in the choice of asymptotic tetrad is henceforth frozen by 
the requirement of a well defined variational principle and Hamiltonian formulation. In contrast, requirement (\ref{bcond}) 
involves additional structure $[\ell]$. The infinitely many possibilities for a metric satisfying Einstein's equations on the 
boundary WIH make it technically very hard to have control of the behavior of the tetrad and connection (along with their 
variations) on $\Delta$ without `gluing them' somehow to the  one available geometric structure $[\ell]$ that remains fixed.

In the treatment in \cite{ENPprl, EngleNouiPerezPranzetti} a complete analysis without this gauge fixing was possible but 
only by restricting to Type I  spherically symmetric  (strongly) isolated horizons with fixed area and a preferred foliation. 
As the discussion in section \ref{S_goodcuts} shows, this effectively singles out a unique geometry on the boundary $\Delta$, 
so that the only allowed variations on $\Delta$ are gauge directions (restricted diffeomorphism and Lorentz gauge transformations) 
of which one does have total control. However, for more general boundary conditions, Type II and III WIHs or even SIHs, 
this does not generalize. 


For general WIHs, we will see that while the Palatini action used in \cite{AFK} is in fact differentiable without any 
gauge fixing or extra boundary terms, the Holst action is not. As it turns out, it is the Holst action and not the Palatini action 
that leads to a canonical formulation in terms of Ashtekar-Barbero variables after a 3+1 decomposition. 
Analogous conclusions are true 
for the self-dual case. As we will clarify in Section \ref{S_decomposition} and show more explicitly in Appendix \ref{A_Spinors}, 
a more restrictive version of the Adapted Null gauge fixing was used in the original treatments \cite{ABF0,ACKclassical, ABF} 
and leads to identical results on differentiability of the self-dual action.

Let us end this part with a remark: Even when the horizon is regarded as an internal boundary, we are still allowing for generic local Lorentz
transformations at the boundary. However, if one were to regard the horizon as a mathematical boundary (in the sense that the region
under consideration is a manifold with boundary), then one would not be allowed to perform transformations that take vectors out of the tangent space
at the boundary. Note that in that case one would be immediately lead to the AN gauge.


\section{Actions}  \label{S_actions}
Historically, a definition for Type I SIHs was coined first and it was spelled out directly in spinorial variables. Taking the  \emph{self-dual action} as a starting point, quantization and derivation of the first law using canonical methods was undertaken in \cite{ABCKprl, ACKclassical,ABKquantum} and \cite{ABF0, ABF} respectively. The covariant phase space accommodating more general  WIHs as an internal boundary of asymptotically flat spacetimes $\mathcal{M}$ and the first law were first derived  from the \emph{Palatini action} in \cite{AFK} and later from the \emph{Holst action}  in \cite{ChatterjeeGhosh}.

Since it is the 3+1 decomposition of the Holst action that leads directly to a canonical formulation in terms of Ashtekar-Barbero variables, and given that we may recover the Palatini and self-dual actions as special cases of the Holst action,
the starting point for our analysis of the variational principle is the covariant and Lorentz-gauge invariant first order \emph{Holst action} \cite{Hojman,holst}: 
 \begin{align}
 S_{\text{Holst}}(e,\A):=&\frac{1}{2\kappa}\,\int_{M}\Sigma^{IJ}\wedge\left(F_{IJ}+\frac{1}{\gamma}\star F_{IJ}\right) \,.  \label{Holst1}
 \end{align}
The two-form
\[
\Sigma^{IJ}:=\star (e^I\wedge e^J)=\frac{1}{2}\epsilon^{IJ}\,_{KL}\,e^K\wedge e^L\,,
\]
is constructed  from the co-tetrad $e_{\mu}^I$, with $\star$ denoting the Hodge dual in the internal space, so that $\epsilon_{IJKL}$ is the Levi-Civita symbol in four dimensions with the convention $\epsilon_{0123}=1$.
The one-form 
\[
{F^I}_J:=\md{\A^I}_J+{\A^I}_K\wedge {\A^K}_J\,,
\]
is the curvature of the Lorentz connection $\A^I_{\mu\,J}$. 
 The arbitrary real or complex constant $\gamma$ denotes the Barbero-Immirzi parameter \cite{Barbero,Immirzi} and as usual $\kappa:=8\pi G$, with $G$ denoting Newton's gravitational constant.

The Palatini action and the self-dual (and anti self-dual) actions can be obtained as special cases of the Holst action (\ref{Holst1}).
The \emph{Palatini action} is obtained as the limit $1/\gamma=0$.
Intuitively, thanks to the isomorphism of Lie algebras $\mathfrak{so}(1,3)\simeq\leftidx{^\pm}{\mathfrak{so}(1,3,\mathbb{C})}{}\simeq\mathfrak{sl}(2,\mathbb{C})$,  where $\leftidx{^\pm}{\mathfrak{so}(1,3,\mathbb{C})}{}$ denote the self- and anti self-dual parts of the complexification of the Lorentz algebra $\mathfrak{so}(1,3,\mathbb{C})\simeq\leftidx{^+}{\mathfrak{so}(1,3,\mathbb{C})}{}\oplus \leftidx{^-}{\mathfrak{so}(1,3,\mathbb{C})}{}$, one may think of internal indices as $\mathfrak{sl}(2,\mathbb{C})$ indices and recover the \emph{self-dual} and \emph{anti-self-dual actions} as the special cases $\gamma=\pm i$. 
More strictly, to recover the self dual action from (\ref{Holst1}) one needs to complexify the tetrad $e^\mu_I$ [and therefore also the NP basis (\ref{spacetimeNPtetrad})], so one may be concerned that imposing reality conditions to select the proper real sector of the complex theory may produce results differing from the real expressions.
We check explicitly in Appendix \ref{A_Spinors} that this is not the case, while making contact with the spinorial $SL(2,\mathbb{C})$ formulation employed in the original treatments on IHs.

\begin{figure}    
     \begin{center}
      \includegraphics[width=10cm]{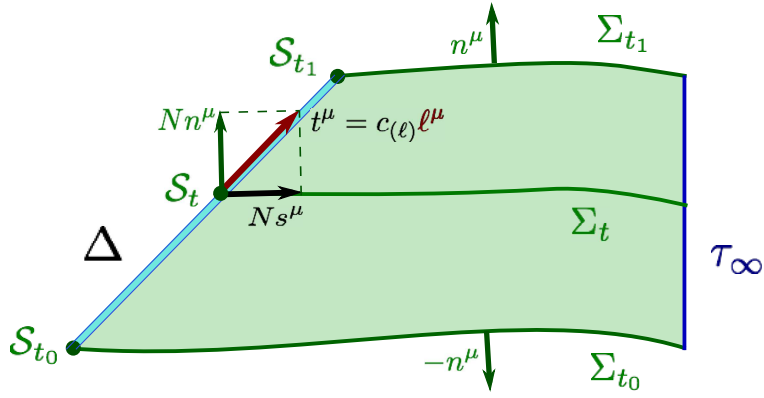}
      \caption{Region of integration $M\subset\mathcal{M}$ with boundary $\partial M=\Sigma_{t_0}\cup\Sigma_{t_1}\cup\Delta\cup\tau_\infty$ and normals. For subsequent 3+1 decomposition, $s^\mu$ is the normal to $\mathcal{S}_t:=\Delta\cap\Sigma_t$ in $\Sigma_t$. For non rotating WIHs, evolution vector $t^\mu$ may be chosen to be proportional to null normal $\ell^\mu$ and equal to $N(n^\mu+s^\mu)$, with $N$ the lapse function (section \ref{S_decomposition}).} \label{fboundaries}
      \end{center}
\end{figure}

The spacetime region of integration $M\subset\mathcal{M}$ for all action principles considered here is bounded by  initial and final spatial Cauchy surfaces $\Sigma_{t_0}$ and $\Sigma_{t_1}$, an internal boundary consisting of  hypersurface $\Delta$ with fixed equivalence class $[\ell]$ (Figure \ref{fWIHgeometry}), and an asymptotic `world tube' $\tau_\infty$ at spatial infinity. In other words $\partial M=\Sigma_{t_0}\cup\Sigma_{t_1}\cup\Delta\cup\tau_\infty$    (Figure \ref{fboundaries}).

For consistency with a 3+1 decomposition, the Cauchy surfaces are required to be asymptotically time translated (cylindrical temporal cut-offs). The boundary $\tau_\infty$ is strictly defined by a limiting process using finite radius $r$ time-like `cylinders' $\mathbb{R}\times S^2_r$ and $r\to\infty$. 
Boundary or fall-off conditions for the fields  $(e^I,\A_{IJ})$ on $\tau_\infty$ are crucial for  finiteness (and differentiability) of the action but are not directly relevant for our analysis on $\Delta$. Details can be found in \cite{CorichiReyes}.

\subsection{Differentiability}

A critical point for the well posedness of a variational principle is the differentiability of the action.
This is understood as the requirement that variations of a first order action $S(e,\A)$ for general relativity must be expressible as
\[
\delta S=\int_M\,\left(\vd{S}{e^I}\right)\delta e^I+\left(\vd{S}{\A_{IJ}}\right)\delta \A_{IJ}\,,
\]
with
\[
\vd{S}{e^I}=0    \qquad \text{ and }  \qquad  \vd{S}{\A_{IJ}}=0
\]
reproducing Einstein's equations. This implies variations of the action have to vanish on the boundary of the region of spacetime considered:  $\delta S|_{\partial M}=0$.
This must be true for a generic point on configuration space and for all variations  of the basic fields $(\delta e^I,\delta \A_{IJ})$ consistent with boundary conditions on $\tau_\infty$ and $\Delta$.
Indeed, while we require fixed configuration fields on the initial and final Cauchy surfaces and hence vanishing variations: $\delta e^I|_{\Sigma_{t_0}}=\delta e^I|_{\Sigma_{t_1}}=\delta \A_{IJ}|_{\Sigma_{t_0}}=\delta \A_{IJ}|_{\Sigma_{t_1}}=0$, on $\tau_\infty$ we must allow variations compatible with asymptotic flatness. Correspondingly, on $\Delta$ we want to allow variations of the possible WIH geometries (\ref{WIHgVar}) which imply compatible nonzero variations of the fields $(\delta e^I,\delta \A_{IJ})|_\Delta$.
For certain applications, we may be content with a fixed geometry on $\Delta$ 
as on calculations of BH entropy. 
However, the more general variations on $\tau_\infty$ and $\Delta$ are crucial for the semiclassical approximation of a path integral formulation to make sense (see e.g. \cite{MannMarolf}).  Furthermore, these more general variations are also required in a derivation of the first law from a canonical Hamiltonian arising from the 3+1 decomposition of the action.

It is well known that for configuration spaces admitting asymptotically flat solutions, the Palatini, self-dual and Holst actions 
must be amended with surface integral terms on $\tau_\infty$ in order to ensure their differentiability.
We will consider the  surface integral term  introduced in \cite{CorichiWilsonEwing}:
\[
-\frac{1}{2\kappa}\int_{\tau_\infty}\Sigma^{IJ}\wedge\left(\A_{IJ}+\frac{1}{\gamma}\star \A_{IJ}\right)\,,
\]
 which generalizes terms previously proposed for the self dual-action \cite{AshtekarLectures} and the Palatini action 
 \cite{AshtekarES}. This counter term makes the Holst action differentiable for asymptotically flat configurations, it has 
 the advantage of being finite and --as it was also shown in \cite{CorichiWilsonEwing}-- it leads to a well defined covariant 
 phase space formulation. Moreover, the same term  also results in a well defined Hamiltonian formulation after a 3+1 decomposition 
 of the action, matching the ADM energy-momentum in Ashtekar-Barbero variables \cite{CorichiReyes}.

The question now remains whether a boundary term at $\Delta$ is necessary when one considers spacetimes admitting general WIHs.
In \cite{AFK}, Ashtekar, Fairhurst and Krishnan showed that no addtional term at $\Delta$ is required if one restricts to the 
Palatini action (although they also explicitly restricted configuration space to the Adapted Null gauge). However, the need for 
a surface term was pointed out already in \cite{ABF} for the self-dual action, and in \cite{ChatterjeeGhosh} and \cite{CRGV2016}  
for the general Holst action. It turns out that the same expression used at $\tau_\infty$ for the Holst action can be used at 
$\Delta$ to write a boundary term that leads to a differentiable action and a well defined covariant formulation if one restricts 
to the Adapted Null gauge.

For that reason we shall consider the action with surface term\footnote{Notice overall minus sign as compared to 
\cite{CorichiWilsonEwing}.}
 \begin{align}
 S(e,\A)=&\frac{1}{2\kappa}\left[\,\int_{M}\Sigma^{IJ}\wedge\left(F_{IJ}+\frac{1}{\gamma}\star F_{IJ}\right) 
 -\int_{\partial{M}}\Sigma^{IJ}\wedge\left(\A_{IJ}+\frac{1}{\gamma}\star \A_{IJ}\right) \,\right]\,.  \label{HolstS}
 \end{align}
A possible reasoning for including the same boundary term at both boundaries is that one could combine them and 
rewrite them as a bulk term. A priory it is not clear whether such strategy shall work, since the details of the boundary 
conditions at infinity and on an internal IH boundary are very different. 
As previously noted, for the Adapted Null gauge such strategy indeed yields a consistent action principle. 
Unfortunately, as we shall verify here, this action is not differentiable if we allow for the full Lorentz gauge freedom at the horizon.
The reason can be directly traced back to the fact that generally the surface term is not Lorentz gauge invariant.

To prove our statements and facilitate comparison with previous treatments, we will expand all fields in terms of the fixed internal null basis $(k^I,l^I,m^I,\bar{m}^I)$ and the corresponding spacetime null basis (\ref{spacetimeNPtetrad}) defined by it. We may then use the Newman-Penrose formalism \cite{NewmanPenrose} when convenient. For conventions the reader is referred to appendix \ref{A_NP}.
In order to take full advantage of null expansions and the Newman-Penrose formalism, one should use a null basis adapted to $\Delta$ as in section \ref{S_preliminaries}, i.e. the null basis element $l^\mu$ should be normal to $\Delta$. 
The simplifications achieved by using a null basis adapted to $\Delta$ are the technical reason behind the restriction to the Adapted Null gauge boundary condition (\ref{bcond}) imposed in previous treatments.
We will generally not require this, since as we have clarified in section \ref{S_gauge}, such choice does partially fix the gauge.  
We consider the full or unrestricted configuration space which is larger from the outset.

For an arbitrary point on the extended configuration space the orthonormal tetrad is expanded as
\[
e_\mu^I=-k_\mu l^I- l_\mu k^I+\bar{m}_\mu m^I+m_\mu\bar{m}^I\,,
\]
and
\begin{align}
\Sigma^{IJ}=\,&2i\big(\,l^{[I}k^{J]}\, m\wedge\bar{m} + l^{[I}\bar{m}^{J]}\, k\wedge m -l^{[I} m^{J]}\, k\wedge\bar{m} \notag\\
&-k^{[I}\bar{m}^{J]}\,l\wedge m + k^{[I}m^{J]}\, l\wedge\bar{m}+ m^{[I}\bar{m}^{J]}\,l\wedge k \,\big)\,.   \label{SigmaIJ}
\end{align}
Expansions above are completely general, independent of any boundary conditions or equations of motion.
On $\Delta$ however, the WIH boundary conditions tell us that Einstein's equation hold.
In particular on $\Delta$, $\A_{IJ}$ matches the spin-connection potential annihilating the tetrad: $\leftidx{^\A}{\mathcal{D}}{_\mu} e^\nu_I=0$. One can use this boundary condition to express $\A_{IJ}$ in terms of certain one-forms determined by the Levi-Civita connection, and correspondingly expand
\begin{align}
\A_{IJ}+\frac{1}{\gamma}\star\A_{IJ}\WIHeq 2\bigg[
&-\left(\W+\frac{1}{\gamma}i\V\right) l_{[I}k_{J]} 
+\left(\bar{\U}+\frac{1}{\gamma}i\bar{\U}\right) l_{[I}m_{J]}  \notag\\
&+\left(\U-\frac{1}{\gamma}i\U\right) l_{[I}\bar{m}_{J]} 
+\left(\V+\frac{1}{\gamma}i\W\right) m_{[I}\bar{m}_{J]}    \notag\\
&+\left(\bar{\Y}-\frac{1}{\gamma}i\bar{\Y}\right) k_{[I}m_{J]} 
+\left(\Y+\frac{1}{\gamma}i\Y\right) k_{[I}\bar{m}_{J]} \,\bigg]\,,  \label{AstarA}
\end{align}
where $\W_\mu:=-k_\nu\nabla_\mu l^\nu$ is a real one-form, $\V_\mu:=\bar{m}_\nu\nabla_\mu m^\nu$ is purely imaginary, and $\U_\mu:=m_\nu\nabla_\mu k^\nu$ and $\Y_\mu:=m_\nu\nabla_\mu l^\nu$ are generally complex.
These one-forms are further related to the Levi-Civita connection $\nabla_\mu$ by equations (\ref{nablaNP1})-(\ref{nablaNP2}) and may be expressed in terms of the Newman-Penrose spin coefficients as in (\ref{WVUYdef2i})-(\ref{WVUYdef2f}). Expansion (\ref{AstarA}) is valid in general  (not only for a point in the Adapted Null gauge) but it requires the half on shell condition $\leftidx{^\A}{\mathcal{D}}{_\mu} e^\nu_I=0$.

We first clarify how the Adapted Null gauge simplifies these expansions. In the Adapted Null gauge $l^\mu=\ell^\mu$ for some $\ell^\mu\in[\ell]$, $im\wedge\bar{m}$ matches the invariant transverse area 2-form (\ref{area2form}): $im\wedge\bar{m}=\leftidx{^2}{\epsilon}{}$,  the pullback of $\W$ matches the pullback of the invariant rotation 1-form (\ref{omegaDef}):  $\W_\pbi{\mu}=\omega_\pbi{\mu}$, and the pullback of $\V$ matches the transverse connection potential (\ref{tV}): $\V_\pbi{\mu}=V_\pbi{\mu}$. Now $\ell_\pbi{\mu}=0$ and it also follows from (\ref{rotationForm}) that $\Y_\pbi{\mu}=0$. So for example, when contracted and pulled back to $\Delta$, only the first term in (\ref{SigmaIJ}) and the first term in (\ref{AstarA}) survive (details are shown in appendix \ref{A_NP}).
In particular, the proposed surface term at $\Delta$ in action (\ref{HolstS}) becomes
\begin{align}
-\frac{1}{2\kappa}\int_\Delta\,\Sigma^{IJ}\wedge\left(\A_{IJ}+\frac{1}{\gamma}\star\A_{IJ}\right)=
&-\frac{1}{2\kappa}\int_\Delta\,2\,im\wedge\bar{m}\wedge\left(\W+\frac{1}{\gamma}i\V\right)  \notag\\
=&-\frac{1}{2\kappa}\int_\Delta\,2\,\leftidx{^2}{\epsilon}{}\wedge\left(\omega+\frac{1}{\gamma}iV\right)  \notag\\
=&-\frac{1}{2\kappa}\int_\Delta\,2\left[\kappa_{(\ell)}+\frac{1}{\gamma}i\,\V\cdot\ell\right]\leftidx{^\Delta}{\epsilon}{}\,,   \label{AFKboundary}
\end{align}
where $\leftidx{^\Delta}{\epsilon}{}:=-ik\wedge m\wedge\bar{m}$ is the  preferred alternating tensor on $\Delta$ and also $\V\cdot\ell=\epsilon_{\text{NP}}-\bar{\epsilon}_{\text{NP}}$, with $\epsilon_{\text{NP}}$ the Newman-Penrose coefficient defined in (\ref{NPcoefs}).

We notice in passing that while the complete surface term is certainly not invariant under general Lorentz transformations, the Palatini part in (\ref{AFKboundary}) is invariant under the restricted $\mathbb{R}^+_\text{global}\rtimes ISO(2)$ transformations preserving the Adapted Null gauge.
However, whereas $im\wedge\bar{m}$ and $\W_\pbi{\mu}$ are invariant under (global) type I and type II and III gauge transformations,
$\V_\pbi{\mu}$ is invariant under all but  type II $U(1)$-rotations, transforming as $\V\to\V+i\md\theta$.
Equivalently, $\epsilon_{\text{NP}}$ transforms as $\epsilon_{\text{NP}}\to\epsilon_{\text{NP}}+\frac{i}{2}\ell^\mu\nabla_\mu\theta$ under type II rotations. So to guarantee invariance of the Holst part of the surface term, one must partially restrict the local  $U(1)$-freedom to type II transformations such that
$\ell^\mu\nabla_\mu\theta=0$, reducing the (already partially gauge fixed) configuration or phase space accordingly.
Finally, for the self-dual case $\gamma=i$, the last expression in (\ref{AFKboundary}) shows that this surface term is a generalization of the one proposed in \cite{ABF}. In that work, for a 3+1 decomposition of the self-dual action rendering a preferred foliation of $\Delta$, the different but related gauge fixing condition $\int_{\mathcal{S}_t}(\V\cdot\ell)\leftidx{^2}{\epsilon}{}=0$ was imposed for every cross sectional two-sphere $\mathcal{S}_t$ on $\Delta$ (in addition to the Adapted Null gauge). This condition makes the last term in (\ref{AFKboundary}) vanish, so the remaining nonzero part matches the term proposed in \cite{ABF}.

We now turn to differentiability of the Holst action with and without a boundary term at $\Delta$. In either case, a surface term at $\Delta$ results from the variation of the corresponding expression for the action. 
One must then check what  the conditions are for each of these terms to vanish.

{\it Action without Boundary Term}. If one does not initially include a boundary term at $\Delta$ in the action  (as in the original treatment \cite{AFK} with 
the Palatini action) then 
\begin{equation} \label{HolstVar1}
\delta S|_\Delta=\frac{1}{2\kappa}\int_\Delta\Sigma^{IJ}\wedge\,\delta\left(\A_{IJ}+\frac{1}{\gamma}\star\A_{IJ}\right)\,.
\end{equation}
We proceed now to refine and extend the arguments in \cite{CRGV2016} that show  this term is generically nonzero, so the 
Holst action without a boundary term at $\Delta$ is not differentiable. We verify that the Palatini part does indeed vanish 
--extending the proof in \cite{AFK} to our larger phase space-- but we show the Holst contribution does not. However, as we 
will see in what follows, it turns out that the Holst action is differentiable in an appropriately 
restricted configuration space. 

The first observation is that term (\ref{HolstVar1}) is gauge invariant. This is most evident from the equivalent expression      
\[
\delta S|_\Delta=\frac{1}{2\kappa}\int_\Delta\left(\Sigma^{IJ}+\frac{1}{\gamma}\star\Sigma^{IJ}\right)\wedge\,\delta\A_{IJ}\,,
\]
given that  $\A_{IJ}$ is a connection and hence $\delta\A_{IJ}$ is a tensor (with respect to internal indices), and so are $\Sigma^{IJ}$ and $\star\Sigma^{IJ}$. 

Let us now restrict to a point $(e^I, \A_{IJ})$ in the Adapted Null gauge and consider first variations along directions which preserve this gauge condition. As with (\ref{AFKboundary}), it is straight forward to verify that in this case surface term (\ref{HolstVar1}) becomes
\begin{equation} \label{HolstVarAFK1}
\delta S|_\Delta=\frac{1}{2\kappa}\int_\Delta2\,\leftidx{^2}{\epsilon}{}\wedge\left(\delta\omega+\frac{1}{\gamma}i\,\delta V\right)\,. 
\end{equation}
Now due to gauge invariance of (\ref{HolstVar1}), it must be true that this last expression is valid for arbitrary points on phase space not necessarily in the Adapted Null gauge, and even if we allow variations along directions not preserving the gauge\footnote{For the committed reader performing calculations, this result might seem surprising at first or even confusing. After all, even for a point in the Adapted Null gauge, variations of (\ref{AstarA}) along directions not preserving this gauge generate more terms. For example, for a variation along a pure type IV gauge rotation: $\delta \W=\bar{\U}\delta b+ \U\delta\bar{b}$, $\delta \V=\bar{\U}\delta b- \U\delta\bar{b}$ and $\delta\Y=\md\delta b-(\W+\V)\delta b$ (see Appendix \ref{A_NP}) . To understand how such terms combine and do not contribute to the variation at the end, one may observe that  an arbitrary variation may be decomposed as the sum of pure variations of the geometry (\ref{WIHgVar})  and variations along pure gauge directions. Pure variations of the geometry are independent of gauge and should therefore also preserve any gauge fixing condition. So to compute variations along such directions at an arbitrary point in configuration space, one may choose a (perhaps different) point representing the same geometry but in the Adapted Null gauge. These variations must therefore match (\ref{HolstVarAFK1}). On the other hand, in the next subsection (and in Appendix \ref{A_NP}), we will show that variations along pure gauge directions actually vanish.}. This gauge invariance is most explicit in the first term. The area two-form is gauge invariant, and since the equivalence class $[\ell]$ is fixed, variations $\delta\omega_\pbi{\mu}$ are solely due to variations of the Levi-Civita connection or ultimately of the spacetime metric.
They do not depend on variations of the NP null basis.  A similar case can be made for $\delta V_\pbi{\mu}$. Since $V_\pbi{\mu}$ represents a $U(1)$-connection,  $\delta V_\pbi{\mu}$ is gauge invariant.

Consider now choosing an $\ell^\mu\in[\ell]$ at each phase space point by some criteria (like fixing the value of $\kappa_{(\ell)}$ as in the derivation of the first law). Whatever the criteria, when we consider a general variation at a fixed point of phase space we must have variations of $\ell^\mu$ be proportional to $\ell^\mu$: $\delta\ell^\mu=\ell^\mu\delta c$ (notice this is different from $\delta l^\mu$ which may not be proportional to $l^\mu$ even at a point in the Adapted Null gauge: $l^\mu\propto\ell^\mu$).
Now at each point of phase space the WIH boundary conditions imply  $\mathcal{L}_\ell\,\omega_\pbi{\mu}\WIHeq 0$. Taking the variation of this equation
then results in:
\begin{equation} \label{Liedw}
 \mathcal{L}_\ell\,\delta\omega_\pbi{\mu}\WIHeq 0\,.
 \end{equation}
Because of our boundary conditions $\delta\omega|_{\Sigma_{t_0}}=0$, this implies $\delta\omega=0$ on all $\Delta$.
So the Palatini contribution in  (\ref{HolstVarAFK1}) vanishes. Notice again that by gauge invariance, (\ref{HolstVarAFK1}) 
is valid for a point on phase space in an arbitrary gauge. Since (\ref{Liedw}) is also clearly independent of any gauge fixing, 
the conclusion holds on all our unrestricted phase space. So the Palatini action (the limit $1/\gamma=0$) is differentiable 
without any gauge fixing.

On the other hand, the additional contribution proportional to $\delta V$ does not vanish. The reason may be related to the fact that the 
Lie derivative of $V$ along $\ell^\mu$ does not generically vanish. Specifically, it can be shown \cite{ChatterjeeGhosh} that $\md V$ is proportional to 
the transverse area form $\leftidx{^2}{\epsilon}{}$ and hence orthogonal to any $\ell\in[\ell]$, so  
$\mathcal{L}_\ell\,V=\md(V\cdot\ell)+\ell\cdot\md V=\md(V\cdot\ell)$. 
Taking the variation of this equation and using again the condition $\delta\ell^\mu=\ell^\mu\delta c$ one 
gets\footnote{Let us note that in \cite{FreidelPerez} there is a proposal for a covariant Lie derivative (that perserves
the covariance under gauge transformations). In particular, the covariant Lie derivative of an $U(1)$ connection is defined as
$L_{\ell}V=\ell\cdot\md V$, and in our case $L_{\ell}V=0$. Nevertheless, $\delta V$ is an $U(1)$ scalar, so its ordinary and
covariant Lie derivatives coincide, so that our condition on $\delta V$ remains the same  in both approaches. }
\begin{equation}
{\cal L}_{\ell}(\delta V) = \delta ( {\cal L}_{\ell}V)-(\delta c)\,{\cal L}_{\ell}V= \md (\ell\cdot\delta V)\, .\label{der-lie}
\end{equation}
This is generically nonzero, which confirms that condition at the boundary $\delta V|_{\Sigma_{t_0}}=0$ is not preserved along $\ell$ unless one restricts to  variations satisfying
\begin{equation}\label{condV}
\ell\cdot\delta V = {\rm const}\, . 
\end{equation}
In the full configuration space (without any gauge fixing) that includes general WIHs or even only general SIHs, there are directions not satisfying the (gauge invariant) condition (\ref{condV}) and correspondingly making a nonzero contribution to the surface integral (\ref{HolstVarAFK1}).
Notice this type of nonzero variations correspond to true variations of the geometry (\ref{WIHgVar}) and exclude variations 
along pure gauge directions. As we show in the next subsection, pure gauge variations actually vanish.
One therefore concludes that for our configuration space of asymptotically flat spacetimes admitting WIHs (or only SIHs) 
as internal boundaries, the Holst action is not differentiable.

To end this part, let us note that one can achieve the vanishing of  (\ref{HolstVarAFK1}), and therefore differentiability of the action, if one restricts the configuration space to points satisfying the stronger requirement:
\begin{equation} \label{rAFKgauge1}
{\cal L}_{\ell} V =\md(V\cdot\ell)=0\, .
\end{equation}
Since this condition is no longer gauge invariant, it necessarily implies a partial gauge fixing, but it certainly does not restrict the space of physical solutions.
For any point in configuration space with given transverse connection potential, if  $V$ does not satisfy (\ref{rAFKgauge1}), by performing a $U(1)$-transformation on the transverse bundle, one can always find another representative $\widetilde{V}=V+\md\Theta_0$  which does so with  $\Theta_0$ solving $\ell\cdot\md\Theta_0+{V}\cdot\ell=$constant.

Preserving condition (\ref{rAFKgauge1}) further restricts $U(1)$-transformations of the transverse connection potential $V\to V+\md\Theta$ to those satisfying
$\ell\cdot\md\Theta=\text{constant}$, so the question remains how this restriction affects the gauge transformations of the action.
For points in the Adapted Null gauge, it is clear that type II gauge transformations acting on the tetrad $e^\mu_I$ and connection $\A^I_{\mu J}$ are in ono-to-one correspondence with $U(1)$-transformations of the transverse connection potental, so (\ref{rAFKgauge1}) imposes the further gauge reduction
\begin{equation}  \label{rU1gauge}
\ell^\mu\nabla_\mu\theta=\text{constant}.
\end{equation}
For points in an arbitrary gauge however, the correspondence is less obvious, in principle the angle $\Theta$ for the $U(1)$-rotation of the transverse connection $V$ could be a complicated function of the Lorentz gauge transformation parameters $\xi$, $\theta$, $a$ and $b$.
Nevertheless,  by performing an explicit calculation, one could convince one-self that this is actually not the case\footnote{For example, one may notice that any point in the extended configuration space may be related to a point in the Adapted Null gauge by performing a type III and a type IV gauge transformation. Writing any point as a point in the Adapted Null gauge rotated successively by a type IV and a type III, one could explicitly perform the variation and verify that $\delta V$ is unaffected.}. 
In this scenario the gauge reduction would be very mild, corresponding only to the partial gauge fixing (\ref{rU1gauge}) of the $U(1)$ gauge rotations of type II along the direction of $\ell$. However, regardless of this very weak gauge reduction (it would not restrict for example the $SU(2)$ gauge after a 3+1 decomposition), a restriction of the configuration space to points satisfying (\ref{rAFKgauge1}) is necessary to ensure differentiability. In the full configuration space the Holst action is not differentiable.

As we also verify explicitly in Appendix \ref{A_Spinors}, expression (\ref{HolstVarAFK1}) with $\gamma=\pm i$ 
matches the self- (and anti self-) dual case. Thus the previous arguments and their conclusion also apply to these 
cases. Without a surface term at $\Delta$, the self (and anti self) dual actions are not differentiable even 
if one restricts to points in the Adapted Null gauge.

{\it Action with Boundary Term}. 
Let us now consider the action with a boundary term at $\Delta$. Varying (\ref{HolstS}) gives
\begin{equation} \label{HolstVar2}
\delta S|_\Delta=-\frac{1}{2\kappa}\int_\Delta\,\delta\Sigma^{IJ}\wedge\left(\A_{IJ}+\frac{1}{\gamma}\star\A_{IJ}\right)\,.
\end{equation}
If one restricts to points in the Adapted Null gauge and to variations preserving the gauge
\begin{equation} \label{HolstVarAFK2}
\delta S|_\Delta=-\frac{1}{2\kappa}\int_\Delta\,2\,\delta\,\leftidx{^2}{\epsilon}{}\wedge\left(\omega+\frac{1}{\gamma}iV\right)\,. 
\end{equation}
A similar argument as used for the Palatini action follows. On a WIH, $\mathcal{L}_\ell\,\leftidx{^2}{\epsilon}{}=0$ and the Adapted Null gauge condition $\delta\ell^\mu=\ell^\mu\delta c$  lead to  $\mathcal{L}_\ell\,\delta\,\leftidx{^2}{\epsilon}{}=0$.
Since variations on the boundary $\partial\Delta$ are required to vanish, this implies again $\delta\,\leftidx{^2}{\epsilon}{}=0$ on all $\Delta$.
Notice however that --unlike (\ref{HolstVar1})-- variation (\ref{HolstVar2}) is not Lorentz gauge invariant, so expression (\ref{HolstVarAFK2}) is only valid for points in the Adapted Null gauge and variations preserving this gauge.
Thus the Holst action (\ref{HolstS}) with surface term at $\Delta$ is differentiable provided one restricts the configuration space to points in the Adapted Null gauge.
This is true for the self-dual case too.

Finally, we turn to differentiability of the action (\ref{HolstS}) but without any gauge fixing.
Since the surface term and hence the whole action is not Lorentz gauge invariant, its variations along pure Lorentz gauge directions cannot vanish in general. In particular the surface term (\ref{HolstVar2}) will generally be nonzero for a pure gauge variation.
This must be true even after one imposes the WIH (or stronger SIH) boundary conditions on $\Delta$ since as we have argued WIH conditions do not involve any (internal) gauge fixing. 
Therefore (\ref{HolstS}) cannot be differentiable in the more general case.
To convince one's self of the non differentiability of the action, one can do explicit calculations. The simplest thing to do is to take a point in configuration space which is in the Adapted Null gauge. Since we already know that surface term  (\ref{AFKboundary}) is not gauge invariant with respect to type IV gauge transformations, we can take a pure type IV gauge variation:
\begin{align*}
\delta l=\bar{m}\,\delta b + m\,\delta\bar{b}, \qquad   \delta{k}=0, \qquad   \delta{m}=k\,\delta b\,.
\end{align*}
Given that we are considering a point in the Adapted Null gauge, we still have $\W_\pbi{\mu}=\omega_\pbi{\mu}$, $\V_\pbi{\mu}=V_\pbi{\mu}$, $l_\pbi{\mu}=0$ and $\Y_\pbi{\mu}=0$, but  now there are more terms in $\delta\Sigma_{IJ}$ and  all terms contribute in (\ref{AstarA}), except of course for the last two. After some algebra we get
\begin{equation}  \label{HolstVarTypeIV}
\delta S|_\Delta=\frac{1}{\kappa}\int_\Delta\left(i-\frac{1}{\gamma}\right)\md(k\wedge\bar{m})\delta b
-\left(i+\frac{1}{\gamma}\right)\md(k\wedge m)\delta \bar{b}\,,
\end{equation}
where we used
$\md(k\wedge \bar{m})\WIHeq(2\alpha-\pi)k\wedge m\wedge\bar{m}$, valid at points in the Adapted Null gauge with forms pulled back to $\Delta$ (for details see Appendix \ref{A_NP}). 
Differentiability hence requires $\md(k\wedge m)=0$, or in terms of Newman-Penrose coefficients 
\[
\pi=2\alpha\,,
\]
but as expected,
this condition is completely gauge dependent. For example, $\pi$ transforms homogeneously under gauge transformations, in particular it is invariant under type I re-scalings  and transforms as $\pi\to e^{-i\theta}\pi$ under type II U(1) rotations, while $\alpha$ transforms as $\alpha\to\alpha+\bar{m}^\mu\nabla_\mu\xi$ under re-scalings and as $\alpha\to e^{-i\theta}(\alpha+i\bar{m}^\mu\nabla_\mu\theta)$ under U(1) rotations.

This requirement is true for general $\gamma$, in particular for the Palatini limit $1/\gamma=0$ and the self- and anti self-dual cases $\gamma=\pm i$, so neither of Holst, Palatini or self-dual actions is differentiable with corresponding surface term.

Notice that the key point in showing differentiability in the Adapted Null gauge of both Palatini without surface term at $\Delta$ and Holst with surface term, is to express the varied surface term integral in terms of variations of quantities which are `time independent' on $\Delta$. Thus, although we are allowing different WIH geometries (\ref{WIHdata}) on our configuration space, the condition $(\delta e^I,\delta \A_{IJ})|_{\Sigma_{t_0}}=0$ effectively sets to zero some of the variations  (\ref{WIHgVar}) of the geometry  at the corresponding $\Delta$ of a given point in configuration space, although not all of them.  As already pointed out in \cite{ABF}, the complete freedom or directions on configuration or phase space can only be `seen' or probed by the Hamiltonian formulation.


\subsection{Type I spherically symmetric SIH case}

As we have already discussed, if one restricts the configuration space to Type I spherically symmetric (strongly) isolated horizons with preferred foliation into good cuts, there is just a two parameter $(a_\Delta,\kappa_{(\ell)})$ family of permissible geometries on $\Delta$. If, as done in \cite{ENPprl, EngleNouiPerezPranzetti}, one fixes the area $a_\Delta$ and the normalization of $\ell^\mu$ so that $\kappa_{(\ell)}=1/R_\Delta$, with aerial radius defined by $a_\Delta=:4\pi R^2_\Delta$, permissible variations are then merely reduced to  pure gauge transformations, namely (internal) Lorentz transformations and diffeomorphisms preserving $\Delta$ (and its foliation). The technical complication of having infinitely many directions to move to, corresponding to different geometries on $\Delta$ compatible with horizon boundary conditions, is circumvented. In order to show differentiability of the Holst action one only needs to consider variations along pure gauge directions. In this very special case,  no surface counter term at $\Delta$ will be required and one does not need to impose any gauge fixing.

Since Holst, as well as Palatini and self-dual actions are gauge invariant (without surface terms at $\Delta$), variations along gauge directions must vanish.
This of course does not directly prove these actions are differentiable when phase space is restricted to Type I SIHs with fixed area and foliations into good cuts because one has to show (\ref{HolstVar1}) by itself vanishes. Nevertheless, it turns out that because of the fixed boundary conditions at $\partial\Delta=\mathcal{S}_{t_0}\cup\mathcal{S}_{t_1}$ this is actually the case.

The proof of differentiability given in \cite{EngleNouiPerezPranzetti} for the self-dual action may be extended to the Holst action. Let us therefore go back and write down the full variation of the Holst action without the boundary at $\Delta$. Since we already know the variation term at $\tau_\infty$ vanishes we do not write it here:
\begin{align}
\delta S=&\int_M -e^J\wedge\left({\epsilon_{IJ}}^{KL}F_{KL}-\frac{2}{\gamma}F_{IJ}\right)\wedge\delta e^I
-\md_\A\left(\Sigma^{IJ}-\frac{1}{\gamma}e^I\wedge e^J\right)\wedge\delta \A_{IJ}  \notag\\
&+\int_\Delta \left(\Sigma^{IJ}-\frac{1}{\gamma}e^I\wedge e^J\right)\wedge\delta \A_{IJ}   \notag
\end{align}
Here $\md_\A$ denotes the covariant exterior derivative, e.g.:
\[
\md_\A\Sigma^{IJ}=\md\Sigma^{IJ}+{\A^I}_K\wedge\Sigma^{KJ}+{\A^J}_K\wedge\Sigma^{IK}\,.
\]
From this variation, a form of Einstein's equations can be read off from the bulk terms.

A general infinitesimal Lorentz gauge transformation may be written as:
\[
\delta{\A^I}_J=-\md(\varepsilon{\tau^I}_J)-[\A,\varepsilon\tau]^I_{\;J}=-\md_\A(\varepsilon{\tau^I}_J)
\]
with ${\tau^I}_J\in\mathfrak{so}(1,3)$ and infinitesimal real parameter $\varepsilon<<1$ (so that ${\Lambda^I}_J=\exp{\varepsilon{\tau^I}_J}$).

Substituting this particular form for the variation $\delta \A_{IJ}$ and integrating by parts, one can show surface term vanishes:
\begin{align}
\int_\Delta \left(\Sigma^{IJ}-\frac{1}{\gamma}e^I\wedge e^J\right)\wedge\delta \A_{IJ}
=&\int_\Delta \md_\A\left(\Sigma^{IJ}-\frac{1}{\gamma}e^I\wedge e^J\right)(\varepsilon\tau_{IJ})  \notag \\
&-\int_{\partial\Delta} \left(\Sigma^{IJ}-\frac{1}{\gamma}e^I\wedge e^J\right)(\varepsilon\tau_{IJ})\,.  \label{LVarG}
\end{align}
First term on the right hand side is zero by Einstein's equations and second term vanishes because $\varepsilon\tau_{IJ}=0$ on $\partial\Delta$.

Similarly for infinitesimal diffeomorphism generators\footnote{Notice these diffeomorphisms need not preserve the preferred foliation.} $X\in T\Delta$:
\[
\delta{\A^I}_J=\mathcal{L}_X{\A^I}_J=X\cdot {F^I}_J+\md_\A(X\cdot{\A^I}_J).
\]
Substituting and integrating by parts the contribution from the second term
\begin{align*}
\int_\Delta \left(\Sigma^{IJ}-\frac{1}{\gamma}e^I\wedge e^J\right)\wedge\delta \A_{IJ}
=&\int_\Delta \left(\Sigma^{IJ}-\frac{1}{\gamma}e^I\wedge e^J\right)\wedge(X\cdot F_{IJ}) \\
 &+\int_\Delta \md_\A\left(\Sigma^{IJ}-\frac{1}{\gamma}e^I\wedge e^J\right)(X\cdot F_{IJ})   \\
&-\int_{\partial\Delta} \left(\Sigma^{IJ}-\frac{1}{\gamma}e^I\wedge e^J\right)(X\cdot F_{IJ})\,.
\end{align*}
Again, the first two terms vanish by Einstein's equations, while the last one is zero because $X=0$ on $\partial\Delta$.

\subsection{Summary}

Let us summarize the results we have found in this part, concerning the differentiability of the actions we have considered.
We can organize these results based on the action and the generality of the configuration or phase space on which this action 
is defined. For the configuration space, we have considered its two aspects: the allowed geometry on the boundary $\Delta$, 
i.e. the type of isolated horizon, and the internal Lorentz gauge symmetry of the field configurations on it, i.e. arbitrary 
versus gauge fixed. 

Let us first recall our findings in the case of the most general WIH boundary conditions. We have showed that, 
the Holst action can be made differentiable without a boundary term at $\Delta$ only in the appropriately restricted configuration
space, where conditions (\ref{rAFKgauge1}) and (\ref{rU1gauge}) are satisfied. 
However, those conditions are sufficient but may not be necessary, and therefore, 
we do not have a complete characterization of the subspace of configuration
space where the action is differentiable. On the full configuration space, the Holst action without a boundary term is not differentiable.

The boundary term given in (\ref{HolstS}) makes the action differentiable in the restricted configuration space with points in 
the Adapted Null gauge, but the amended action is still non-differentiable in the larger configuration space where there is 
no gauge fixing. 
These results are valid for every $\gamma$, in particular for $\gamma =i$, that is the self-dual action
with or without the boundary term. Also, conclusions above do not change if we restrict the allowed geometries at $\Delta$ 
to general SIHs. On the other hand, the Palatini action (the limit $1/\gamma=0$) is differentiable without the need of a 
surface term at $\Delta$ in the larger configuration space without gauge fixing. The boundary term makes it non-differentiable 
in the general case. It remains differentiable only in the Adapted Null gauge. 

Finally, restricting the configuration space to Type I strongly isolated horizons with preferred foliation the situation is quite 
different. In this case, the only allowed field variations on the horizon are Lorentz gauge transformations and diffeomorphisms 
compatible with the boundary conditions. It turns out that the self-dual, Palatini and Holst actions without boundary terms at 
$\Delta$ are differentiable.  This is not the case when the corresponding boundary term at $\Delta$ is included. Again this is 
due to the non gauge invariance of the surface terms considered.
Our analysis is summarized in Table \ref{difTable}.


\begin{table}[tb]
\begin{tabular}{| c || c | c || c |}
\hline
           & \multicolumn{2}{c||}{General WIHs or SIHs} & Type I SIHs \\  \cline{2-3}
Action & No gauge  & Adapted Null & with   \\
 & fixing  & gauge &   preferred foliation \\
\hline
Holst & $\times$ & $\times$ &  $\surd$  \\
\hline
Holst+BT &  $\times$ & $\surd$ &  $\times$ \\
\hline
Palatini & $\surd$ & $\surd$ &   $\surd$ \\
\hline
Palatini+BT & $\times$ & $\surd$ &  $\times$  \\  
\hline
Self-dual & $\times$ & $\times$ &  $\surd$ \\
\hline
Self-dual+BT & $\times$ & $\surd$ &  $\times$  \\
\hline
\end{tabular}
\caption{Differentiability for first order actions with and without  boundary term (BT) at $\Delta$.}
\label{difTable}
\end{table}


\


\section{3+1 decomposition, foliations and residual gauge}  \label{S_decomposition}

In this section we shall analyse the interplay between the Adapted Null gauge and 3+1 decomposition of the action.
We will henceforth restrict configuration space to points in the Adapted Null gauge.
A full 3+1 analysis will be carried out elsewhere \cite{CRVwih2}. 	   

As it is usual for a 3+1 decomposition and Hamiltonian formulation of a covariant theory, one postulates a time function $t:M\subseteq\mathcal{M}\to\mathbb{R}$ on (a portion of) spacetime $\mathcal{M}$, whose level curves $\Sigma_t$ are spatial hypersurfaces and provide a foliation of spacetime region $M$. Spacetime fields are split into tangential (spatial) and normal components with respect to this foliation.  For the configuration space consisting of spacetimes admitting a weakly isolated horizon $\Delta$ as an internal boundary, we will only consider foliations of the bulk of $M$ whose intersection with $\Delta$ is non trivial and induces a foliation of $\Delta$ by two-spheres $\mathcal{S}_t=\Sigma_t\cap\Delta$.

Additionally, one needs to choose an \emph{evolution} vector field $t^\mu$ such that $t^\mu\nabla_\mu t=1$ and along which spatial 
fields are defined to `evolve'.  The evolution vector field will generically be time-like on the bulk and (for non rotating horizons) 
it will be chosen to belong to the equivalence class $[\ell]$ on $\Delta$. The latter boundary choice makes $t^\mu$ coincide with 
symmetry directions on $\Delta$ so that the Hamiltonian or canonical generator of diffeomorphisms along $t^\mu$ on the boundary will 
correspond to a conserved quantity or \emph{energy} associated to the WIH $\Delta$.




On $\Delta$ the boundary condition for evolution vector field $t^\mu$ is then
\begin{equation} \label{evolutionField}
t^\mu=\ct\ell^\mu
\end{equation}
where as before $\ell^\mu:=e^\mu_Il^I \in[\ell]$ for the Adapted Null gauge, and $\ct$ is a constant on a given spacetime (point of phase space). 
To be consistent with results of covariant phase space formulation, in particular with the first law, $\ct$ must 
be a  function of geometric invariant $a_\Delta$, which changes from point to point in phase space.\footnote{Recall that in the  covariant phase space formulation, the energy of the horizon $E^t_\Delta$, as
defined by the time translation $t^\mu$, satisfies a first law type relation
$\delta E^t_\Delta =  \frac{1}{8\pi G}\k_{(t)}\de a_\Delta$. Thus, a necessary and sufficient condition for the existence of the horizon energy $E^t_\Delta$ is that $\k_{(t)}$ can only depend on $a_\Delta$. 
The first law  is a condition on the choice of the vector field $t^\mu$, such that
the vector field $\de_t$ on covariant phase space $\Gamma$ is Hamiltonian. Each permissible $t^\mu$ defines a horizon energy $E^t_\Delta$. 
One can select a preferred vector field $t^\mu$ among permissible evolution vector fields that will lead to a definition of mass of the isolated horizon \citep{AFK}.}
This selects a particular representative from $[\ell]$ by requiring $\kappa_{(t)}$ to have 
a specific functional dependence on $a_\Delta$. 
In particular one can choose
\[
\kappa_{(t)}:=\frac{1}{2R_\Delta}\, , 
\]
where $a_\Delta=4\pi R_\Delta^2$. In this way on a static solution $t^\mu$ coincides with the Killing vector field on $\Delta$.
Then, for this choice of $t^\mu$ we can define the horizon mass as $M_\Delta = E^t_\Delta$, for any point of the covariant
phase space, not only for static spacetimes. 

To find $\ct$ we notice that it must be true that
\[
\kappa_{(t)}=\ct\kappa_{(\ell)}
\]
so that
\[
\ct=\frac{\kappa_{(t)}}{\kappa_{(\ell)}}=\frac{\kappa_{(t)}}{\omega\cdot\ell}\,.
\]
Notice $\ct$ changes as $\ct\to\xi^{-1}\ct$ under constant rescalings $\ell\to\xi\ell$.


\subsection{Compatibility with the 3+1 foliation}

Decomposition for the spacetime metric splits the ten independent components of $g_{\mu\nu}$ into the six independent components of the Euclidean spatial metric $q_{\mu\nu}$, the lapse function $N$ and the  shift vector $N^\mu$. Lapse and shift being respectively the normal and tangential components of the evolution vector field $t^\mu=Nn^\mu+N^\mu$, where again, $n^\mu$ denotes the future directed normal to the foliation and $N=-\left|(\md t)_\mu(\md t)^\mu\right|^{-1/2}$. In coordinates $(t,y^a)$ adapted to the foliation the line element reads
\begin{equation} \label{ADMmetric}
g_{\mu\nu}\md x^\mu \md x^\nu=(-N^2+q_{ab}N^aN^b)\md t^2+2q_{ab}N^b\md t\,\md y^a + q_{ab}\md y^a \,\md y^b\,.
\end{equation}
Since $t^\mu$ is also a normal on $\Delta$, these adapted coordinates are \emph{comoving with} $\Delta$.

\begin{figure}    
     \begin{center}
      \includegraphics[width=7.2cm]{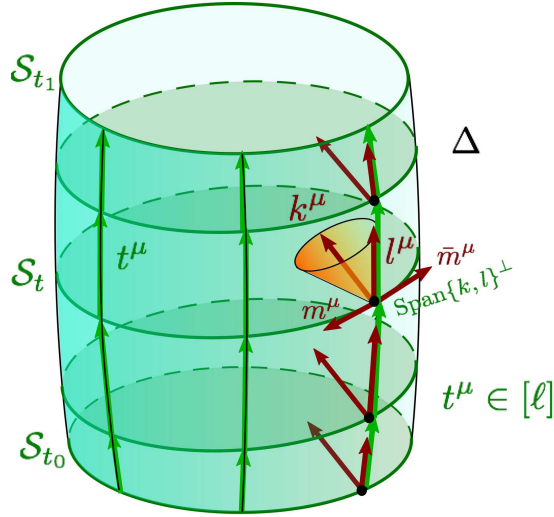}
      \caption{A foliation $\{\Sigma_t\}$ for $M$ and Lie dragged by $t^\mu$ is compatible with $(\Delta, [\ell])$ if it defines a foliation $\{\mathcal{S}_t=\Sigma_t\cap\Delta\}$ by 2-spheres for $\Delta$ and also $t^\mu\in [\ell]$. In the Comoving gauge, the Newman-Penrose null basis $(k^\mu,l^\mu,m^\mu,\bar{m}^\mu)$ defined by isomorphism (\ref{spacetimeNPtetrad}) is such that both $k^\mu$ and $l^\mu$ are normal to the foliation $\{\mathcal{S}_t\}$ with $l^\mu\in[\ell]$. Equivalently, $l^\mu\in[\ell]$ and Span$\{k,l\}^\bot$ defines an integrable distribution tangent to the foliation.} \label{fWIHfoliation}
      \end{center}
\end{figure}

The \emph{compatibility} condition $t^\mu\nabla_\mu t=1$ and the choice of evolution vector field (\ref{evolutionField}) on $\Delta$ imply $\ell^\mu$ corresponds to the velocity field of the  null generators $\gamma^\mu(\lambda)$ parameterized by $\lambda=c_{(t)}t$. Given this condition, there is still plenty of freedom in the choice of foliation.  This is because in order to construct such \emph{compatible foliations}\footnote{Notice this definition is slightly different to compatible foliations as defined in \cite{GJreview} because ours are  foliations compatible with (Lie dragged by) $t^\mu=c_{(\ell)}\ell^\mu$ instead of $\ell^\mu$. Again, this is a choice of normalization that we have made to be consistent with the first law.} one may choose different  `initial' surfaces $\Sigma_{t_0}$ (determining different transverse spheres $\mathcal{S}_{t_0}$ on $\Delta$) to be Lie dragged along $t^\mu$ (Fig. \ref{fWIHfoliation}). On $\Delta$, this freedom amounts to a `time translation' $t \to t+T$ reparameterization at each null generator, with function $T$ constant along each generator, i.e. $\mathcal{L}_{{\ell}}T=0$.

As we have already discussed, a given  foliation also selects a particular transverse null direction $\mathbf{k}^\mu$ from the infinitely many choices in the null cone at each point $p\in\Delta$. This is the unique null direction normal to $\mathcal{S}_t$ apart from $\ell^\mu$ and such that $\mathbf{k}_\mu\ell^\mu=-1$.  
Conversely, given $\mathbf{k}^\mu$ such that the orthogonal planes $\text{Span}\{\mathbf{k},\ell\}^\bot\subset T_pM$ at each point $p\in\Delta$ define an integrable distribution, this determines a foliation.
Notice we now use a different notation for this transverse null direction field $\mathbf{k}^\mu$ determined by a given foliation.
This is to emphasize that, a priori, it need not be related or coincide with the transverse null direction determined by the tetrad (\ref{spacetimeNPtetrad}): $k^\mu:=e^\mu_Ik^I$ (or any other arbitrary transverse null direction).

One can spell out the relation between $\mathbf{k}^\mu$ and the 3+1 foliation more explicitly. As already stated, a foliation in the bulk is assumed to induce a foliation of the boundary by two spheres $\Delta\approx[t_0,t_1]\times\mathcal{S}_t$. Let $s^\mu$ be the `outward pointing' unit normal to $\mathcal{S}_t$ in $\Sigma_t$ (Figure \ref{fboundaries}), i.e.
\[
s_\mu s^\mu=1,   \qquad  s_\mu n^\mu=0,  \qquad \text{ and } \qquad   s_\mu v^\mu=0,   \quad \text{ for } v^\mu\in T_p\mathcal{S}_t.
\]
Since $T_p\mathcal{S}_t\subset T_p\Sigma_t$ and $T_p\mathcal{S}_t\subset T_p\Delta$, it also follows that $n_\mu v^\mu=0$ and $\ell_\mu v^\mu=0$, and hence
\[
0=\ct\ell_\mu v^\mu=t_\mu v^\mu=(Nn_\mu+N_\mu)v^\mu=N_\mu v^\mu.
\]
So the shift vector must be proportional to the unit normal to the two-spheres:  $N^\mu=\alpha s^\mu$, and from the condition $\ell_\mu\ell^\mu=0$, it follows (assuming $N>0$ for future directed normals $\ell^\mu$ and $n^\mu$) that
\begin{equation}  \label{ellequation}
\ell^\mu=\frac{N}{\ct}\left(n^\mu+s^\mu\right)\,.
\end{equation}
The other `ingoing' null direction normal to $\mathcal{S}_t$ must then be proportional to $n^\mu-s^\mu$, so choosing the normalization condition $\mathbf{k}\cdot\ell=-1$ then
\begin{equation}  \label{kequation}
\mathbf{k}^\mu=\frac{\ct}{2N}\left(n^\mu-s^\mu\right)\,.
\end{equation} 
Inverting these relations we get
\begin{equation} \label{nequation}
n^\mu=\frac{\ct}{2N}\ell^\mu+\frac{N}{\ct}\mathbf{k}^\mu\,.
\end{equation}
Explicitly a differential equation determining $(\md t)_\mu$.
Solving for $s^\mu$ in (\ref{ellequation}) and substituting in (\ref{kequation}), one may also see that a transverse field $\mathbf{k}^\mu$ determined by a foliation satisfies $\mathbf{k}_\pbi{\mu}=-c_{(\ell)}(\md t)_\pbi{\mu}$.
A change of foliation corresponding to reparameterization $t \to t+T$ then changes $\mathbf{k}^\mu$ as  $\mathbf{k}_\mu\to \mathbf{k}_\mu-c_{(\ell)}(\md T)_\mu$.

Thus given a foliation, the transversal direction $\mathbf{k}^\mu$ is completely determined by (\ref{kequation}) modulo constant rescalings or global type I Lorentz transformations (manifested in $\ct$). 
If one now \emph{requires} -- just as one did for $l^\mu$ -- that the transverse null direction defined by the isomorphism induced by the tetrad (\ref{spacetimeNPtetrad}) coincides with  this $\mathbf{k}^\mu$, i.e. $k^\mu:=e^\mu_Ik^I=\mathbf{k}^\mu$,
then clearly this compatibility requirement implies a further gauge reduction to $\mathbb{R}^+_{\text{global}}\rtimes U(1)$. Indeed, to preserve the Adapted Null gauge and compatibility with the foliation, both directions $k^\mu$ and $l^\mu$ have to remain fixed under gauge transformations of the tetrad $e^\mu_I\to \Lambda_I^{\;\;J}\,e^\mu_J$, so only gauge transformations  amounting to global rescalings of type I and $U(1)$-rotations of type II on the null tetrad defined by (\ref{spacetimeNPtetrad}) are allowed.
We may call this compatibility condition the \emph{Comoving gauge} (Fig. \ref{fWIHfoliation}).
The Comoving gauge was used in the initial treatments on isolated horizons \cite{ABF, ACKclassical}. In fact, it was essentially imposed from the very beginning as part of the first definitions of isolated horizons which relied on a preferred foliation for $\Delta$.

\subsection{Time gauge compatibility}

Independently of compatibility with the foliation or Comoving gauge, a further reduction of the gauge group $\mathbb{R}^+_{\text{global}}\rtimes ISO(2)$ on $\Delta$ is forced on us by the time gauge. The time gauge fixing is indispensable to derive the Hamiltonian formulation of general relativity in Ashtekar-Barbero variables, directly from the 3+1 decomposition of the Holst action.

Let $n^I:=e^I_\mu n^\mu$ be the (inverse) image of the spacetime normal to the foliation under the isomorphism induced by the tetrad. The \emph{time gauge} consists in choosing an orthonormal basis in internal Minkowski space such that
\begin{equation} \label{internalNormal}
n^I=\eta^{IJ}n_J=\delta^I_0\,,
\end{equation}
or in component form $n^I=(1,0,0,0)$ and $n_I=(-1,0,0,0)$. 
Since this implies $n^\mu=e^\mu_In^I=e^\mu_I\delta^I_0=e_0^\mu$, the time gauge is equivalent to first choosing an arbitrary orthonormal basis for the internal space and then restricting to those tetrads for which 
\begin{equation}  \label{timeGauge}
e_0^\mu=n^\mu\,,
\end{equation}
so the remainig vectors $e^\mu_i$, with $i=1,2,3$, are purely spatial. This condition restricts Lorentz gauge transformations on $M$ to those preserving (\ref{timeGauge}), that is to $SO(3)$ rotations.
 
So, starting with the Holst action (\ref{HolstS}) with configuration space restricted to the Adapted Null gauge, performing a 3+1 decomposition and subsequently imposing the time gauge,  necessarily implies a gauge reduction to $U(1)$. The reason being that the full compact group $SO(3)$ cannot be a subgroup of $\mathbb{R}^+\rtimes ISO(2)$. The only subgroups of $SO(3)$ preserving a null vector are necessarily isomorphic to $U(1)$ or the identity. 
 
The final question remaining then is what is the most general $\mathbb{R}^+\rtimes ISO(2)$-gauge reduction  --as determined by the covariant analysis-- compatible with the 3+1 foliation and the time gauge.
To answer it, we first choose an arbitrary orthonormal basis in the internal space and let $\ell^I=\left(\ell^0, \ell^i\,\right)$. From (\ref{ellequation}) and the time gauge (\ref{timeGauge}) it follows that
\[
\ell^\mu=\ell^Ie^\mu_I=\ell^0e_0^\mu+\ell^ie_i^\mu=\ell^0n^\mu+\ell^ie^\mu_i=\ct^{-1}N\left(n^\mu+s^\mu\right)\,,
\]
and from the orthogonality of the triad and the normal vectors we have in particular
\begin{equation}  \label{fixedLapse}
\ell^0=\frac{N}{\ct}\,.
\end{equation}
Using this relation and (\ref{spacetimeNPtetrad}) with the Comoving gauge, equation (\ref{nequation}) implies
\[
n^I=\frac{1}{2\ell^0}\ell^I+\ell^0k^I\,,
\]
or equivalently 
\[
k^I=\left(k^0,k^i\,\right)=\left(\frac{1}{2\ell^0},-\frac{\ell^i}{2(\ell^0)^2}\right)\,,
\]
which shows that given an arbitrary internal $\ell^I=\left(\ell^0, \ell^i\,\right)$, compatibility of the Comoving gauge with the time gauge completely fixes $k^I$. Additionally, we have 
\[
m^I=(0,m^i)\,,
\]
with the corresponding orthogonality and normalization conditions
\[
\ell_i\ell^i=(\ell^0)^2, \quad  m_im^i=0, \quad m_i\ell^i=0,  \quad \text{ and } \quad  m_i\bar{m}^i=1\,.
\]

Finally, equation (\ref{fixedLapse}) also shows how the time gauge further reduces the Comoving $\mathbb{R}^+\rtimes U(1)$ gauge imposed by the foliation, to simply $U(1)$. Indeed, the requirement that the lapse $N$ is gauge invariant freezes out the rescaling freedom in $\ct$ and hence global transformations of type I or rescalings of $\mathbf{k}$ and $\ell$.
However, we emphasize here again that this particular reduction only follows if one requires that the null normal to $\Delta$  and the transverse direction associated to the foliation coincide with those elements of the null spacetime tetrad (\ref{spacetimeNPtetrad}), 
i.e. if one is in the Comoving gauge.


\section{Outlook}   \label{S_outlook}

In this manuscript we have revisited the issue of internal gauge invariance, within the first order formulation of general relativity, for weakly isolated horizons. This issue becomes relevant when one is interested in obtaining a Hamiltonian description for the gravitational degrees of freedom in the presence 
of a horizon as a boundary of the region under consideration. The first step in such endeavour is to have a well defined variational principle consistent with the boundary conditions on the geometrical variables. 
While the definition of a horizon is purely geometrical and in principle does not entail a restriction on the variables used to describe gravity --be it metric or tetrad variables--, one has to analyse in detail the corresponding action and its gauge symmetries, in order to have a consistent description. 

In this contribution, we have argued that for the Holst action principle that contains the Palatini and self-dual actions as special cases, and yields a simple canonical 3+1 description of gravity,  one can obtain a consistent action principle when: i) One includes a boundary term for the horizon and, ii) One restricts the internal gauge freedom for the first order variables. Furthermore, if one wants to have a canonical 3+1 description of the theory, as is needed in canonical approaches to quantum gravity, then the gauge freedom is further reduced to a U(1) theory on the horizon. 
As noted in the main part, one could attain a differentiable action principle for the Holst action without a boundary term but that entails a reduction of the configurations space that is not properly understood. This issue deserves further attention. One should also note that, in the 3+1 canonical decomposition, one may obtain a consistent Hamiltonian description in this case. Those findings shall be reported elsewhere \cite{CRVwih2}. 
Finally, the special case of Type I horizons, that posses more symmetry needs to be treated separately and in that case, an SU(2) formulation is possible. 

More precisely, we have shown that the proposed boundary terms in the literature for the Holst (and the self-dual) action necessitate a gauge reduction of the covariant theory to $\mathbb{R}^+_{\text{global}}\rtimes ISO(2)$ in order to guarantee differentiability of the action accommodating WIHs as internal boundaries. This fact  inevitably implies a further gauge reduction to $U(1)$ in the Hamiltonian theory on the horizon boundary. This however does not rule out the possibility of constructing an appropriate boundary term (in the covariant or Hamiltonian theory) which makes the Holst or Hamiltonian action differentiable and leads to an effective $SU(2)$ theory on a general isolated horizon.  
There have been other proposals for boundary terms in the vielbein formalism that are gauge invariant 
\cite{Obukhov, BianchiWieland, Bodendorfer2013}. 
However, those terms involve expressions that can only be understood on null surfaces via a limiting process. It would 
be worthwhile to extend the definition of those terms for null boundaries. Furthermore, those terms are ill-defined for 
asymptotically flat boundary conditions.
Thus, we see that the available boundary terms do not have what is required of them, namely to be well defined on null boundaries {\em and} yield a consistent action principle for asymptotically flat configurations. The burden is on finding a term that satisfies such conditions.


\begin{acknowledgments}
This work was in part supported by DGAPA-UNAM IN103610 grant, by CONACyT 0177840 
and 0232902 grants, by the PASPA-DGAPA program, by NSF
PHY-1403943 and PHY-1205388 grants,  by the Eberly Research Funds of Penn State,
and by CIC, UMSNH.
\end{acknowledgments}


\appendix


\section{Expansions in arbitrary Newman-Penrose null basis}  \label{A_NP}

Regardless of any null hypersurface one may always introduce  a (non coordinate) Newman-Penrose null basis $(k^\mu,l^\mu,m^\mu,\bar{m}^\mu)$. This is certainly most convenient when dealing with general null hypersurfaces  and  in particular it has proved very useful  when working with weakly isolated horizons $\Delta$, but as we have emphasized in the main text, the NP bases defined by (\ref{spacetimeNPtetrad}) need not be adapted to $\Delta$. In this appendix we show some details of the expansions in terms of these bases.

For the Newman-Penrose formalism we follow the same conventions of \cite{AFK}. In particular the Newman-Penrose spin coefficients are defined as
\begin{align}
\kappa_{\text{NP}}&:=-m^\mu l^\nu  \nabla_\nu l_\mu   &\epsilon_{\text{NP}}:=2^{-1}(\bar{m}^\mu  l^\nu \nabla_\nu m_\mu -k^\mu  l^\nu \nabla_\nu l_\mu) \qquad \pi&:=\bar{m}^\mu l^\nu\nabla_\nu k_\mu  \notag\\
\sigma&:=-m^\mu m^\nu  \nabla_\nu  l_\mu    &\beta:=2^{-1}(\bar{m}^\mu m^\nu  \nabla_\nu m_\mu -k^\mu m^\nu \nabla_\nu l_\mu)   \qquad   \mu&:=\bar{m}^\mu m^\nu\nabla_\nu k_\mu\notag\\
\rho&:=-m^\mu \bar{m}^\nu \nabla_\nu  l_\mu     &\alpha:=2^{-1}(\bar{m}^\mu \bar{m}^\nu  \nabla_\nu m_\mu -k^\mu \bar{m}^\nu \nabla_\nu l_\mu)    \qquad \lambda&:=\bar{m}^\mu\bar{m}^\nu\nabla_\nu k_\mu  \label{NPcoefs} \\
\tau&:=-m^\mu k^\nu \nabla_\nu  l_\mu   &\gamma_{\text{NP}}:=2^{-1}(\bar{m}^\mu k^\nu  \nabla_\nu m_\mu -k^\mu k^\nu \nabla_\nu l_\mu)
 \qquad  \nu&:=\bar{m}^\mu k^\nu\nabla_\nu k_\mu        \notag 
\end{align}

From the Levi-Civita connection one may also define one-forms
\begin{equation} \label{WVUYdef}
\W_\mu:=-k_\nu\nabla_\mu l^\nu\,, \qquad \V_\mu:=\bar{m}_\nu\nabla_\mu m^\nu\,, \qquad \U_\mu:=m_\nu\nabla_\mu k^\nu\,, \qquad \Y_\mu:=m_\nu\nabla_\mu l^\nu\,.
\end{equation}
Using orthogonality, one can directly expand
\begin{align}
\nabla_\mu l^\nu=&\W_\mu l^\nu+\bar{\Y}_\mu m^\nu+\Y_\mu\bar{m}^\nu\,,  \label{nablaNP1}\\
\nabla_\mu k^\nu=&-\W_\mu k^\nu+\bar{\U}_\mu m^\nu+\U_\mu\bar{m}^\nu\,,  \label{nablaNPk} \\
\nabla_\mu m^\nu=& \U_\mu l^\nu+\Y_\mu k^\nu+\V_\mu m^\nu\,.  \ \label{nablaNP2}
\end{align}
From their definition (\ref{WVUYdef}),  it is straight forward to derive how these forms transform under general Lorentz (type I, II, III and IV) transformations of the NP tetrad \cite{BasuChatterjeeGhosh} (see below).
In terms of the NP spin coefficients one also has
\begin{align}
\W_\mu:=& -(\gamma_{\text{NP}}+\bar{\gamma}_{\text{NP}}) l_\mu-(\epsilon_{\text{NP}}+\bar{\epsilon}_{\text{NP}})k_\mu+(\alpha+\bar{\beta})m_\mu+(\bar{\alpha}+\beta)\bar{m}_\mu\,, \label{WVUYdef2i} \\
\U_\mu:=&-\bar{\nu} l_\mu-\bar{\pi}k_\mu+\bar{\mu}m_\mu+\bar{\lambda}\bar{m}_\mu\,, \label{WVUYdef2i2}\\
\V_\mu:=&  -(\gamma_{\text{NP}}-\bar{\gamma}_{\text{NP}}) l_\mu-(\epsilon_{\text{NP}}-\bar{\epsilon}_{\text{NP}})k_\mu+(\alpha-\bar{\beta})m_\mu+(\beta-\bar{\alpha})\bar{m}_\mu\,,   \\
\Y_\mu:=& \tau l_\mu+\kappa_{\text{NP}}k_\mu+\rho m_\mu-\sigma\bar{m}_\mu\,.   \label{WVUYdef2f}
\end{align}

All definitions and expressions above are independent of  any null hypersurface. In the Adapted Null gauge, i.e. if $l^\mu=\ell^\mu$ coincides with normal to null hypersurface $\Delta$, one has
\[
\ell_\pbi{\mu}=0  \qquad    \text{ and }   \qquad  \kappa_{\text{NP}}=0.
\]
If furthermore $\Delta$ is a NEH then
\[
\rho=0    \qquad    \text{ and }   \qquad   \sigma=0.
\]
So $\Y_\mu= \tau\ell_\mu$ and hence $\Y_\pbi{\mu}=0$. By comparing definitions  (\ref{WVUYdef}) with (\ref{omegaDef}) and (\ref{tV}), it is clear that in this gauge $\omega_\pbi{\mu}=\W_\pbi{\mu}$ and $V_\pbi{\mu}=\V_\pbi{\mu}$, and since $\kappa_{(\ell)}=\epsilon_{\text{NP}}+\bar{\epsilon}_{\text{NP}}$, from (\ref{HajicekForm}) we have 
$$\Omega_\mu= -(\gamma_{\text{NP}}+\bar{\gamma}_{\text{NP}})\ell_\mu+(\alpha+\bar{\beta})m_\mu+(\bar{\alpha}+\beta)\bar{m}_\mu.$$

In the Comoving gauge $k^\mu=\mathbf{k}^\mu$ is normal to a foliation of $\Delta$, so $\underleftarrow{k}\propto-\underleftarrow{\md t}$ implies $\underleftarrow{\md k}=0$. Pulling back and anti symmetrizing  (\ref{nablaNPk}) and substituting (\ref{WVUYdef2i}) and (\ref{WVUYdef2i2}) one infers
\[
\pi=\alpha+\bar{\beta} \qquad \text{ and } \qquad \mu=\bar{\mu}\,.
\]

The spin connection potential ${\A^I}_J$ satisfying $\leftidx{^\A}{\mathcal{D}}{_\mu}e^\nu_I=0$, may be expanded in terms of one-forms (\ref{WVUYdef}) [and the fixed internal null tetrad $(k^I, l^I, m^I, \bar{m}^I)$]. Indeed, writing
\[
\A_{IJ}=A\,l_{[I}k_{J]}+B\,l_{[I}m_{J]}+C\,l_{[I}\bar{m}_{J]}+D\,k_{[I}m_{J]}+E\,k_{[I}\bar{m}_{J]}+F\,m_{[I}\bar{m}_{J]}\,,
\]
and comparing (\ref{nablaNP1})-(\ref{nablaNP2}) with
$\nabla_\mu l^\nu=\leftidx{^\A}{\mathcal{D}}{_\mu}(e^\nu_Il^I)=e^\nu_I\leftidx{^\A}{\mathcal{D}}{_\mu}l^I=e^\nu_I\A^I_{\mu J}l^J$,
and similarly $\nabla_\mu k^\nu=e^\nu_I\A^I_{\mu J}k^J$   and  $\nabla_\mu m^\nu=e^\nu_I\A^I_{\mu J}m^J$, one gets equations
\begin{align*}
\A^I_{\mu J}l^J=&\,\W_\mu l^I+\bar{\Y}_\mu m^I+\Y_\mu\bar{m}^I\,,  \\
\A^I_{\mu J}k^J=&-\W_\mu k^I+\bar{\U}_\mu m^I+\U_\mu\bar{m}^I\,, \\
\A^I_{\mu J}m^J=&\, \U_\mu l^I+\Y_\mu k^I+\V_\mu m^I\,,
\end{align*}
from which to solve for the undetermined coefficients:
\begin{equation}  \label{Aexpansion}
\A_{IJ}=-2\W\,l_{[I}k_{J]}+2\bar{\U}\,l_{[I}m_{J]}+2\U\,l_{[I}\bar{m}_{J]}+2\bar{\Y}\,k_{[I}m_{J]}+2\Y\,k_{[I}\bar{m}_{J]}+2\V\,m_{[I}\bar{m}_{J]}\,.
\end{equation}

From this expression (\ref{AstarA}) follows. Contracting the latter with (\ref{SigmaIJ}) one gets
\begin{align}
\Sigma^{IJ}\wedge\left(\A_{IJ}+\frac{1}{\gamma}\star\A_{IJ}\right)\WIHeq 2i\bigg[
&(m\wedge\bar{m})\wedge  \left(\W+\frac{1}{\gamma}i\V\right) 
-(k\wedge m)\wedge           \left(\bar{\Y}-\frac{1}{\gamma}i\bar{\Y}\right)  \notag\\
&+(k\wedge\bar{m})\wedge  \left(\Y+\frac{1}{\gamma}i\Y\right)
+(l\wedge m)\wedge             \left(\bar{\U}+\frac{1}{\gamma}i\bar{\U}\right)  \notag\\
&-(l\wedge\bar{m})\wedge  \left(\U-\frac{1}{\gamma}i\U\right)  
-(l\wedge k)\wedge             \left(\V+\frac{1}{\gamma}i\W\right)   
 \,\bigg]\,.   \label{generalExpansion}
\end{align}
The same expansion follows for (\ref{HolstVar1}) and (\ref{HolstVar2}) except variation $\delta$ acts on the one-forms $\W$, $\V$, $\U$, $\Y$ (and  conjugates) in big parentheses in the first case and on the null basis in smaller parentheses in the second case.
Clearly, in the Adapted Null gauge, when pulled back to $\Delta$ only the first term survives and (\ref{AFKboundary}) follows.

Let us now use the expansion analogous to (\ref{generalExpansion}) corresponding to the variation term (\ref{HolstVar2}) but evaluated at a point in the Adapted Null gauge and pulled back to $\Delta$:
\begin{align}
\delta\Sigma^{IJ}\wedge\left(\A_{IJ}+\frac{1}{\gamma}\star\A_{IJ}\right)\WIHeq 2i\bigg[
&\delta(m\wedge\bar{m})\wedge  \left(\omega+\frac{1}{\gamma}iV\right) 
+(\delta l\wedge m)\wedge             \left(\bar{\U}+\frac{1}{\gamma}i\bar{\U}\right)  \notag\\
&-(\delta l\wedge\bar{m})\wedge  \left(\U-\frac{1}{\gamma}i\U\right)  
-(\delta l\wedge k)\wedge             \left(V+\frac{1}{\gamma}i\omega\right)   
 \,\bigg]\,.   \label{AFKexpansion2}
\end{align}
To avoid cluttering notation, in this expression all forms are understood to be pulled back to $\Delta$ and we have used $\W_\pbi{\mu}=\omega_\pbi{\mu}$, $\V_\pbi{\mu}=V_\pbi{\mu}$, $\Y_\pbi{\mu}=0$ and $l_\pbi{\mu}=0$.
Notice that for variations preserving the Adapted Null gauge condition one has $\delta l=l\,\delta c$ (which is hence zero when pulled back to $\Delta$) and $\delta(im\wedge\bar{m})=\delta\,\leftidx{^2}{\epsilon}{}$, so only the first term in (\ref{AFKexpansion2}) survives and  (\ref{HolstVarAFK2}) follows.

To show explicitly that (\ref{HolstS}) is not differentiable let us consider a variation corresponding to an infinitesimal type IV Lorentz transformation, so
\[
\delta(m\wedge\bar{m})=k\wedge \bar{m}\,\delta b - k\wedge m\,\delta\bar{b}  \qquad \text{ and } \qquad \delta l=\bar{m}\,\delta b+m\,\delta\bar{b}\,.
\]
To simplify (\ref{AFKexpansion2}) with this type of variation, one may use expansions (\ref{WVUYdef2i})-(\ref{WVUYdef2f}) in terms of the Newman-Penrose spin coefficients. This gives in particular
\begin{align*}
\delta(im\wedge\bar{m})\wedge  \left(\omega+\frac{1}{\gamma}iV\right)
\WIHeq&\left[-i(\alpha+\bar{\beta})+\frac{1}{\gamma}(\alpha-\bar{\beta})\right] k\wedge m\wedge\bar{m}\,\delta b \,+\,\text{complex conjugate}\,,   \\
i\delta l\wedge m \wedge \left(\bar{\U}+\frac{1}{\gamma}i\bar{\U}\right)
\WIHeq&\left(i-\frac{1}{\gamma}\right)\pi\,k\wedge m\wedge\bar{m}\,\delta b \,,    \\
-i\delta l\wedge k \wedge             \left(V+\frac{1}{\gamma}i\omega\right) 
\WIHeq&\left[-i(\alpha-\bar{\beta})+\frac{1}{\gamma}(\alpha+\bar{\beta})\right] k\wedge m\wedge\bar{m}\,\delta b \,+\,\text{complex conjugate}\,, 
\end{align*}
where again, forms are understood to be pulled back to $\Delta$.
Adding these three equalities plus the complex conjugate of the second one, one finally gets
\[
\delta\Sigma^{IJ}\wedge\left(\A_{IJ}+\frac{1}{\gamma}\star\A_{IJ}\right)
\WIHeq 2\left(i-\frac{1}{\gamma}\right)\left(\pi-2\alpha\right)k\wedge m\wedge\bar{m}\,\delta b \,+\,\text{complex conjugate}\,,
\]
which gives (\ref{HolstVarTypeIV}).
To write this independently of NP spin coefficients, one can use 
\begin{equation}  \label{dkwedgebm}
\md(k\wedge\bar{m})\WIHeq(2\alpha-\pi)k\wedge m\wedge\bar{m}
\end{equation}
which follows from using (\ref{nablaNPk}) and (\ref{nablaNP2}) in $\md(k\wedge\bar{m})=\md k\wedge\bar{m}-k\wedge\md\bar{m}$,  substituting  (\ref{WVUYdef2i})-(\ref{WVUYdef2f}) and then pulling back to $\Delta$.

As an aside, while we have a general proof showing (\ref{HolstVar1}) vanishes for variations along pure gauge directions, it is instructive to see how this works out directly from expansion\footnote{Both the proof and the expansion use the same on shell boundary condition or Einstein equation, namely the compatibility of the connection with the tetrad: $\leftidx{^\A}{\mathcal{D}}{_\mu}e^\nu_I=0$.}  (\ref{generalExpansion}).
We can see this explicitly at least for points in the Adapted Null gauge and Lorentz gauge transformations.
For this purpose, let us now use the expansion analogous to (\ref{generalExpansion}) corresponding to the variation term (\ref{HolstVar1}), evaluated at a point in the Adapted Null gauge and pulled back to $\Delta$:
\begin{align}
\Sigma^{IJ}\wedge\delta\left(\A_{IJ}+\frac{1}{\gamma}\star\A_{IJ}\right)\WIHeq 2i\bigg[
&(m\wedge\bar{m})\wedge  \left(\delta\W+\frac{1}{\gamma}i\delta\V\right) 
-(k\wedge m)\wedge           \left(\delta\bar{\Y}-\frac{1}{\gamma}i\delta\bar{\Y}\right)  \notag\\
&+(k\wedge\bar{m})\wedge  \left(\delta\Y+\frac{1}{\gamma}i\delta\Y\right)
 \,\bigg]\,.   \label{AFKExpansion1}
\end{align}
Under Lorentz gauge transformations $\W$, $\V$, $\U$ and $\Y$ transform as \cite{BasuChatterjeeGhosh}
\[
\begin{array}{ll}
\text{type I:} & \; \W\to\W+\md\ln\xi,            \quad  \V\to\V,                       \quad  \U \to \xi^{-1} \U,   \quad  \Y\to\xi \Y   \\
\text{type II:} & \;  \W\to\W,                  \quad  \V\to\V+i\md\theta,      \quad  \U\to e^{i\theta}\U,  \quad \Y\to e^{i\theta}\Y  \\
\text{type III:} & \; \W\to\W-\bar{a}\Y-a\bar{\Y},  \; \V\to\V-\bar{a}\Y+a\bar{\Y},  \; \U\to\U-\md a +a(\W-\V)-a^2\bar{\Y}, \;  \Y\to\Y \\
\text{type IV:} & \; \W\to\W+\bar{b}\U+b\bar{\U},  \; \V\to\V-\bar{b}\U+b\bar{\U},  \; \U\to\U,  \;  \Y\to \Y+\md b -b(\W+\V)-b^2\bar{\U}   
\end{array}
\]
From which it follows in particular
\begin{align}
\text{type I:}  \;\;\; &   \quad \delta\W=\md\delta\xi,  \quad     \delta\V=0,  \quad     \delta\Y=\Y\delta\xi  \label{LVarI}\\
\text{type II:} \;\, &   \quad \delta\W=0,      \quad    \delta\V=i\md\delta\theta,  \quad   \delta\Y=i\Y\delta\theta \label{LVarII} \\
\text{type III:} &   \quad \delta\W=-\Y\delta\bar{a}-\bar{\Y}\delta a,   \quad   \delta\V=-\Y\delta\bar{a}+\bar{\Y}\delta a   \quad  \delta\Y=0  \label{LVarIII}\\
\text{type IV:} &   \quad \delta\W=\bar{\U}\delta b+\U\delta\bar{b},   \quad   \delta\V=\bar{\U}\delta b-\U\delta\bar{b},  \quad   \delta\Y=\md \delta b-(\W+\V)\delta b  \label{LVarIV}
\end{align}
Substituting (\ref{LVarI}) in (\ref{AFKExpansion1}) and using $\Y_\pbi{\mu}=0$, one gets a total differential matching the last term on the right hand side of (\ref{LVarG}):
\[
\Sigma^{IJ}\wedge\delta\left(\A_{IJ}+\frac{1}{\gamma}\star\A_{IJ}\right)\WIHeq 2i\,\md(m\wedge\bar{m}\,\delta\xi)\,.
\]
Similarly for (\ref{LVarII}) one gets
\[
\Sigma^{IJ}\wedge\delta\left(\A_{IJ}+\frac{1}{\gamma}\star\A_{IJ}\right)\WIHeq \frac{2i}{\gamma}\md(m\wedge\bar{m}\,\delta\theta)\,.
\]
For (\ref{LVarIII}), variation (\ref{AFKExpansion1}) vanishes identically. Finally, variation (\ref{LVarIV}) is slightly more complicated, but again, substituting (\ref{WVUYdef2i})-(\ref{WVUYdef2f}) and using (\ref{dkwedgebm}) one gets also a total differential:
\[
\Sigma^{IJ}\wedge\delta\left(\A_{IJ}+\frac{1}{\gamma}\star\A_{IJ}\right)\WIHeq 2\left(i-\frac{1}{\gamma}\right)\md(k\wedge\bar{m}\,\delta b)\,+\,\text{complex conjugate}\,.
\]


\section{Spinorial formulation} \label{A_Spinors}

In this appendix we specialize to the self-dual action. We work directly with spinor variables which are best suited for this case. We re-derive key expressions obtained for the more general Holst action in Appendix \ref{A_NP} and  make contact with original treatments \cite{ABF0, ABF, ACKclassical}.

We follow the same conventions used in \cite{ABF} and \cite{AshtekarLectures}.
Now the internal Minkowski space is replaced by internal complex two dimensional spinor space $\mathbb{C}^2$, with
symplectic structure or metric $\epsilon_{AB}$ with inverse $\epsilon^{AB}$ such that  $\epsilon_{AB}=-\epsilon_{BA}$ and  $\epsilon^{AB}\epsilon_{AC}=\delta^B_C$.
$A$, $B$, $C$ are spinor indices running from 1 to 2. 
Just like the metric $\eta_{IJ}$ and its inverse in Minkowski space, $\epsilon_{AB}$ and $\epsilon^{AB}$ are used
 to raise and lower spinor indices:  $X_B:=X^A\epsilon_{AB}$ and $X^A=\epsilon^{AB}X_B$, which implies $X^AY_A=-X_AY^A$.  
Complex conjugation $X^A\mapsto\bar{X}^{A'}$, maps spinors to the \emph{conjugate} spinor space with metric $\bar{\epsilon}_{A'B'}$.

Analogously to fixing a null tetrad $(k^I,l^I,m^I,\bar{m}^I)$ in internal Minkowski space, one fixes a dyad basis $(\om^A,\iota^A)$ in spinor space $\mathbb{C}^2$ such that  
\[
\iota^A\iota_A=0   \qquad\qquad  \om^A\om_A =0  \qquad\qquad \iota^A\om_A=-\iota_A\om^A=1
\]
and similarly for the conjugate basis $(\bar{\om}^{A'},\bar{\iota}^{A'})$. One sets $\bar{\partial}(\om^A,\iota^A)=0$ [and also $(\delta\om^A,\delta\iota^A)=0$].

The orthonormal tetrad $e^\mu_I$ is replaced by the \emph{soldering vector} $\sigma^\mu_{AA'}$ providing an isomorphism between the space of anti-hermitian spinors $X^{AA'}$ such that $\bar{X}^{AA'}=-X^{AA'}$ and the tangent space $T_p\mathcal{M}$ at each point $p$. 
Its inverse the \emph{soldering form} $\sigma_\mu^{AB}$ is obtained by lowering spacetime indices with $g_{\mu\nu}$ and raising spinor indices with $\epsilon^{AB}$ and $\bar{\epsilon}^{A'B'}$.
They satisfy
\[
\sigma_\mu^{AA'}\sigma^\nu_{AA'}=\delta^\nu_\mu,  \qquad  \sigma_\mu^{AA'}\sigma^\mu_{BB'}=\delta^{A}_{B}\delta^{A'}_{B'}
\qquad \text{ and }  \qquad    \bar{\sigma}^\mu_{AA'}=-\sigma^\mu_{AA'}\,.
\]

Analogously to (\ref{spacetimeNPtetrad}), the soldering vector defines a Newman-Penrose null basis:
\begin{align}
l^\mu:=i\sigma^\mu_{AA'}\om^A\bar{\om}^{A'}    \qquad \qquad  &m^\mu:=i\sigma^\mu_{AA'}\om^A\bar{\iota}^{A'}  \notag\\
k^\mu:=i\sigma^\mu_{AA'}\iota^A\bar{\iota}^{A'}  \qquad \qquad  &\bar{m}^\mu:=i\sigma^\mu_{AA'}\iota^A\bar{\om}^{A'} \label{spinorIsomorphism}
\end{align}

The spin connection compatible with the soldering form or vector: $\nabla_\mu\sigma^\nu_{AA'}=0$, acts as
\[
\nabla_\mu X^{AA'}=\bar{\partial}_\mu X^{AA'}+\leftidx{^+}{\A}{^A_{\;\;B}}X^{BA'}+\leftidx{^-}{\bar{\A}}{^{A'}_{\;\;B'}}X^{AB'}
\]
where $\leftidx{^+}{\A}{^A_{\;\;B}}$ is precisely the self-dual connection potential and  
$\leftidx{^-}{\bar{\A}}{^{A'}_{\;\;B'}}$ is the conjugate of the anti self-dual connection potential. 

One can then find an expansion in terms of the one-forms  (\ref{WVUYdef}) and the dyad $(\iota_A,\om_A)$ for the self-dual connection potential,  analogous to expansion (\ref{Aexpansion}) for the real spin connection and corresponding to (\ref{AstarA}).
Writing
\[
\leftidx{^+}{\A}{_{AB}}=2X\,\iota_{(A}\om_{B)}+Y\,\om_A\om_B+Z\,\iota_A\iota_B
\]
and comparing (\ref{nablaNP1})-(\ref{nablaNP2}) with
\[
\nabla_\mu l^\nu=\nabla_\mu(i\sigma^\nu_{AA'}\om^A\bar{\om}^{A'})=i\sigma^\nu_{AA'}\nabla_\mu(\om^A\bar{\om}^{A'})
=i\sigma^\nu_{AA'}(\leftidx{^+}{\A}{^A_{\mu B}}\om^B\bar{\om}^{A'}+\leftidx{^-}{\bar{\A}}{^{A'}_{\mu B'}}\om^A\bar{\om}^{B'})
\]
and similarly 
\[
\nabla_\mu k^\nu=i\sigma^\nu_{AA'}(\leftidx{^+}{\A}{^A_{\mu B}}\iota^B\bar{\iota}^{A'}+\leftidx{^-}{\bar{\A}}{^{A'}_{\mu B'}}\iota^A\bar{\iota}^{B'})
\]
and  
\[
\nabla_\mu m^\nu=i\sigma^\nu_{AA'}(\leftidx{^+}{\A}{^A_{\mu B}}\om^B\bar{\iota}^{A'}+\leftidx{^-}{\bar{\A}}{^{A'}_{\mu B'}}\om^A\bar{\iota}^{B'})\,,
\]
one gets equations
\begin{align*}
\leftidx{^+}{\A}{^A_{\mu B}}\om^B-\leftidx{^-}{\bar{\A}}{^{A'}_{\mu B'}}\om^A\bar{\om}^{B'}\bar{\iota}_{A'}
=&\,\W_\mu \om^A+\Y_\mu \iota^A \\
\leftidx{^+}{\A}{^A_{\mu B}}\iota^B+\leftidx{^-}{\bar{\A}}{^{A'}_{\mu B'}}\iota^A\bar{\iota}^{B'}\bar{\om}_{A'}
=&-\W_\mu \iota^A+\bar{\U}_\mu \om^A  \\
\leftidx{^+}{\A}{^A_{\mu B}}\om^B+\leftidx{^-}{\bar{\A}}{^{A'}_{\mu B'}}\om^A\bar{\iota}^{B'}\bar{\om}_{A'}
=&\, \Y_\mu \iota^A+\V_\mu \om^A
\end{align*}
from which to solve for the undetermined coefficients:
\begin{equation}
\leftidx{^+}{\A}{_{AB}}=-(\W+\V)\,\iota_{(A}\om_{B)}+\bar{\U}\,\om_A\om_B-\Y\,\iota_A\iota_B
\end{equation}

On the other hand, one has
\begin{align}
\Sigma^{AB}&=\sigma^{AA'}\wedge\sigma^B_{\;\;A'}  \notag\\
&=2[\iota^A\iota^Bl\wedge m+\iota^{(A}\om^{B)} (m\wedge\bar{m}-l\wedge k)-\om^A\om^B k\wedge\bar{m}].
\end{align}
Contracting the last two expressions gives:
\[
\Sigma^{AB}\wedge\leftidx{^+}{\A}{_{AB}}=2\,l\wedge m\wedge\bar{\U}+(m\wedge\bar{m}-l\wedge k)\wedge(\W+\V)
+2\,k\wedge\bar{m}\wedge\Y\,,
\]
which as expected, matches (\ref{generalExpansion}) for $\gamma=i$, up to an overall factor of $2i$.
Hence also $\Sigma^{AB}\wedge\,\delta \leftidx{^+}{\A}{_{AB}}$ and $\delta\Sigma^{AB}\wedge\leftidx{^+}{\A}{_{AB}}$ match their corresponding expansions computed in Appendix \ref{A_NP} (regardless of any gauge fixing) and indeed all discussions and conclusions for the Holst action apply to the self-dual case.

In the original treatments \cite{ABF, ACKclassical}, a preferred foliation for $\Delta$ was singled out from the outset and the null vectors $l^\mu$ and $k^\mu$ defined by (\ref{spinorIsomorphism}) were required to be normal to the two spheres $\mathcal{S}_t$.
So in our terminology, the comoving gauge was essentially imposed from the beginning: $(k^\mu\propto\mathbf{k}^\mu, l^\mu\propto\ell^\mu)$. The equivalence class $[\ell]$, or more specifically the null normal $\ell^\mu$ was singled out by the normalization $\kappa_{(\ell)}=1/2R_\Delta$. The (Type I strongly) isolated horizon conditions were given directly in terms of the spin connection compatible with the soldering form and the gauge fixed adapted NP tetrad:
\[
\om^A\nabla_\pbi{\mu}\om_A\WIHeq 0    \qquad \text{and} \qquad   \iota^A\nabla_\pbi{\mu}\iota_A\WIHeq \mu\bar{m}_\pbi{\mu}\,,
\]
with spherically symmetric NP coefficient $\mu$.


\end{document}